\documentclass[fleqn,usenatbib]{mnras}

\usepackage{newtxtext,newtxmath}

\usepackage[T1]{fontenc}

\DeclareRobustCommand{\VAN}[3]{#2}
\let\VANthebibliography\thebibliography
\def\thebibliography{\DeclareRobustCommand{\VAN}[3]{##3}\VANthebibliography}

\usepackage{graphicx}	
\usepackage{amsmath}	
\usepackage[normalem]{ulem}

\newcommand{\ramses}{{\rm \small RAMSES}\ }

\newcommand{\rascas}{{\rm \small RASCAS}\ }

\newcommand{\Msolyr}{{\rm\, M_\odot\,yr^{-1}}}
\newcommand{\Msol}{{\rm\, M_\odot}}

\newcommand{\Halpha}{{\rm H$\alpha$}\ }

\newcommand{\HI}{H{\small I}\ }

\newcommand{\kmsec}{${\,\rm {km\,s^{-1}}}$\ }
\newcommand{\kmsecalt}{${\,\rm {km\,s^{-1}}}$}

\title[]{Forming the local starburst galaxy Haro 11 through hydrodynamical merger simulations}

\author[T. Ejdetjärn et al.]{
Timmy Ejdetjärn$^{1}$\thanks{E-mail:timmy.ejdetjarn@gmail.com},
Oscar Agertz$^{2}$,
Florent Renaud$^{3,4}$,
Göran Östlin$^{1}$,
\newauthor 
\ Alexandra Le Reste$^{5}$,
Angela Adamo$^{1}$
\\\\
$^{1}$Oskar Klein Centre, Department of Astronomy, Stockholm University, 106 91 Stockholm, Sweden\\
$^{2}$Department of Astronomy and Theoretical Physics, Lund Observatory, Box 43, SE-221 00 Lund, Sweden\\
$^{3}$Observatoire Astronomique de Strasbourg, Universit\'e de Strasbourg, CNRS UMR 7550, F-67000 Strasbourg, France\\
$^{4}$University of Strasbourg Institute for Advanced Study, 5 all\'ee du G\'en\'eral Rouvillois, F-67083 Strasbourg, France\\
$^{5}$Minnesota Institute for Astrophysics, School of Physics \& Astronomy, University of Minnesota, 116 Church St. SE, Minneapolis, MN 55455, USA
}

\date{Accepted XXX. Received YYY; in original form ZZZ}

\pubyear{2025}

\begin{document}
\label{firstpage}
\pagerange{\pageref{firstpage}--\pageref{lastpage}}
\maketitle

\begin{abstract}
Haro 11 is a metal-poor, starburst galaxy believed to be the result of an ongoing merger, which is shaping the properties of the galaxy. In this study, we carry out a large suite of numerical simulations of a merger between two disc galaxies, to study possible origins of Haro 11 and understand under which conditions various features of the galaxy are formed. By varying galaxy parameters describing the orbital configurations, masses, and their inclination, we perform a total of $\sim500$ simulations. We demonstrate that a two-disc galaxy merger reproduces key, observed features of Haro 11, including its morphology, gas kinematics, star formation history, and stellar population ages and masses. In particular, we present a fiducial Haro 11 model that produces the single observed tidal tail, three stellar knots, and inner gas morphology and kinematics. The resulting orbit and galactic morphology are robust against small variations of the initial parameters. By performing mock observations, we compare with the results of observational data and discuss possible origins for various features. Furthermore, we present newly gathered observational data that confirms the presence of a stellar tidal tail with similar length and morphology as our simulations.

\end{abstract}



\section{Introduction}
Understanding the interplay of in-situ and external mechanisms on the evolution of galaxies remains a central challenge in extragalactic astronomy. Galaxy interactions and mergers are significant events in the formation history of a galaxy, found to be drivers of starbursts and large-scale morphological transformations \citep[e.g.][]{MihosHernquist1996, Li+08, Lambas+12, Renaud+22}, with possible ramifications for properties such as: massive star cluster formation \citep[e.g.][]{Kruijssen+12, Adamo+20}, AGN fueling \citep[]{Sharma+24, LaMarca+24}, and the escape of ionising radiation (e.g. \citealt{KostyukCiardi2024}; \citealt{Zhu+24}; but see \citealt{Mascia+25} for counter-examples). Thus, knowing the specific scenario of a galaxy merger enables a deeper insight into the observed properties and physical mechanisms at work within interacting galaxies.

Blue compact dwarfs/galaxies (BCD/BCG) are compact and metal-poor galaxies undergoing a period of intense starburst resulting in especially young stellar populations dominating the blue part of the spectrum of the galaxy. Most BCGs have irregular morphologies \citep[e.g.][]{BergvallOstlin2002, Micheva+13}, which would indicate current or recent close interaction and/or merger with a companion \citep[e.g.][]{Ostlin+01, Adamo+11}. This can also be inferred from tidal tails or gas debris, as well as periods of starburst. BCGs exhibit an extreme environment of stellar and cluster formation, and due to their low metallicity, compact size, and high SFR have been suggested to be analogues of young starburst galaxies in the high-$z$ Universe \citep[e.g.][]{Ostlin+01, Adamo+11, Sirressi+22, Gao+22}. 

The Haro 11 galaxy is a local BCG/BCD that has been studied thoroughly in the UV, IR, sub-mm, radio, visible, and X-ray \citep[e.g.][]{Ostlin+99, Bergvall+06, Adamo+10, Ostlin+15, Pardy+16, Gao+22, Danehkar+24, LeReste+24}. At a distance of 88.5 Mpc \citep[][]{Sirressi+22}, it is the closest confirmed Lyman continuum (LyC) leaker \citep[][]{Bergvall+06, Rivera-Thorsen+17, Ostlin+21, Komarova+24} and a bright Ly${\rm \alpha}$ emitter \citep[][]{Hayes+07, Ostlin+09}. This makes Haro 11 a valuable local analogue to understand what physical processes facilitate the escape of ionising radiation in high-$z$ systems. Furthermore, its rich observational dataset makes Haro 11 particularly well-suited for simulations aimed at understanding the physical processes that shape such systems.


The stellar population and recent starburst \citep[e.g.][]{Adamo+10, Sirressi+22, Chandar+23}, complex gas kinematics \citep[e.g.][]{Ostlin+01, Ostlin+15, Menacho+21, Gao+22}, and irregular stellar and gas morphology \citep[e.g.][]{Adamo+10, Ostlin+15, Menacho+21}, all suggest that Haro 11 is the result of an ongoing merger. Indeed, it was only recently that the presence of a tidal tail was observed in \HI emission from 21 cm observations by \citet{LeReste+24}, which serves as a strong indication of an ongoing merger. Applying these observations as constraints, numerical models of galaxy mergers offer an avenue to confirm if these features are consistent with specific merger scenarios and explain the origin of various other properties of Haro 11.

\begin{figure*}
    \includegraphics[width=0.48\textwidth]{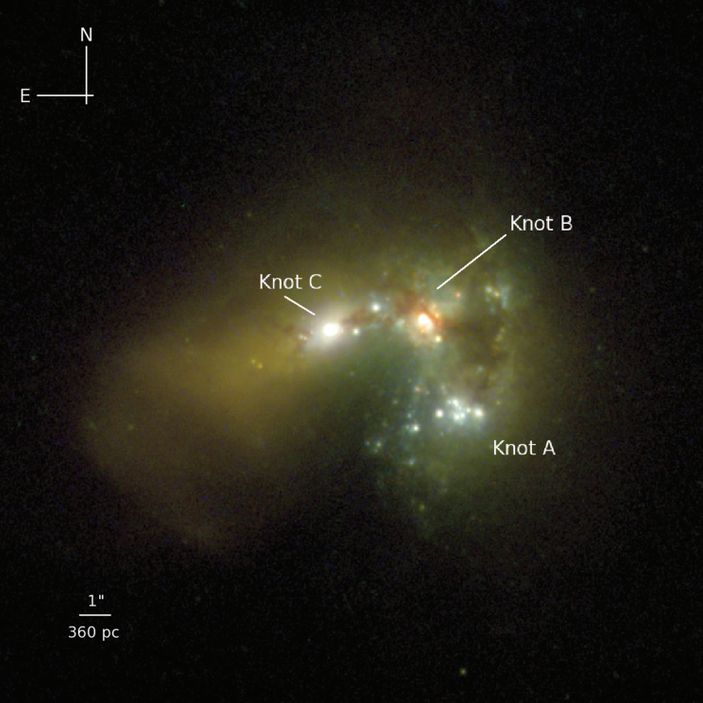}
    \includegraphics[width=0.48\textwidth]{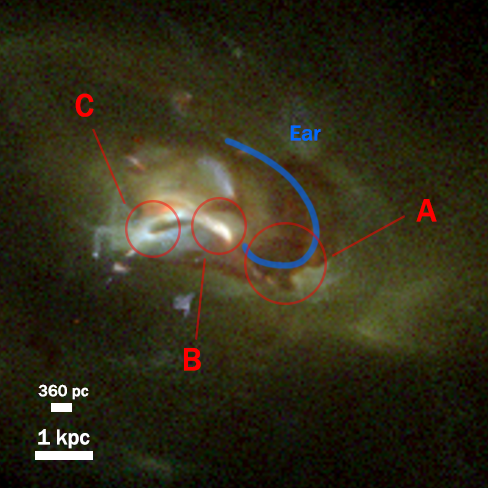}
    \caption{\emph{Left:} Image of Haro 11 from \citet{Adamo+10}, taken by HST with the waveband filters F220W, F435W, and F814W. \emph{Right:} Mock observations of the simulated galaxy, applying the same HST filters. Details of how this mock image was produced is described in Section~\ref{sec:visualise_morphology}. Both images encompass a $\sim8$\,kpc box and have a resolution of $\sim 30-40$\,pc. }
    \label{fig:mock}
\end{figure*}

To model a specific (observed) galaxy merger, we require observational constraints that map to orbital parameters. In particular, tidal tail features and stellar properties are strong constraints, but with some degeneracy between parameters. However, previous numerical work of galaxy interactions have managed to pinpoint the initial orbital parameters of the progenitor galaxies that reproduce the general morphology and inherent galaxy properties. For example, simulations of the Antennae galaxies have managed to reproduce the morphology of the galaxies and their tidal tails, as well as the star formation, among other properties \citep{Karl+10, Privon+13, Renaud+15, Lahen+18}. In \citet[][hereafter Ö15]{Ostlin+15}, the authors provided a comprehensive comparison between the Haro 11 galaxy and the Antennae galaxies. The Antennae galaxies are undergoing a merger and the authors show it to have similar morphology and kinematics as Haro 11. Although the Antennae galaxies are more massive and have two tidal tails, the fact that they have comparable kinematics and morphology suggests their orbital parameters might be similar.

However, high-resolution simulations of specific galaxy mergers are still very limited. Even more rare are simulations of BCG/BCD \citep[][]{Bekki2008, Pearson+18, Zhang+20, Leicester+24}, which are useful resources to interpret observational data and deepen our understanding about the complex physical mechanisms going on in these extreme galaxy environments. However, finding initial conditions that show a broad agreement across different observed features is not trivial and requires an extensive study of the parameters controlling the galaxy orbits.  

In this paper, we model the merger of two disc galaxies to reproduce the general morphology and properties of the Haro 11 galaxy as observed today, and discuss a possible origin of its formation. We employ the hydrodynamical and $N$-body code \ramses \citep[][]{Teyssier2002} to perform a suite of low-resolution simulation tests ($\sim 100$\,pc) of the interaction, varying the galaxy and orbital parameters until the merging galaxies' morphology resembles the morphology of Haro 11 today. We compare with observational studies of the gaseous and stellar components, e.g. star formation rate and masses, to constrain the initial conditions our model. The simulation that best represents Haro 11 observational properties is then run at the higher spatial resolution of 36 pc. We present the results of this fiducial simulation and compare the successes and shortcomings of our models and discuss its applications. Furthermore, we present newly acquired deep imaging data that show the presence of a stellar tidal feature in the direction of the \HI gas tail which have been predicted by the fiducial model, corroborating the good match.

The paper is structured as follows. Section \ref{sec:obs_constraint} details observations of Haro 11 that are useful for constraining the initial conditions of the progenitor galaxies and their orbital parameters. In Section \ref{sec:simulations} we detail our numerical approach, as well as our methodology to achieve the best match with observational data by varying orbital and galactic parameters. In Section~\ref{sec:haro11_model} we present our fiducial model and directly compare its inner and tidal tail morphology, star formation rate, stellar population properties, and kinematics with observational data. Section~\ref{sec:extra_tests} summarises the results of this parameters study, highlighting parameters that are important for producing specific galaxy properties. We then discuss the agreements and mismatches between our fiducial model and observation in Section~\ref{sec:discussion}. Finally, we summarise and discuss our results in Section~\ref{sec:conclusion}.

\section{Observational constraints}\label{sec:obs_constraint}
In this section, we briefly outline the key components of Haro 11 that were used to constrain and confirm the initial properties and orbital parameters of the progenitor galaxies. All of these observational constraints were considered during the formulation of our simulations' initial conditions, but not all of them were used, as matching every property in detail is not feasible. The method of determining our initial conditions is explained in Section~\ref{sec:ICs}. 

\subsection{Morphological appearance}
A distinct feature of Haro 11 is its three bright stellar knots A, B, and C (see left plot of Figure~\ref{fig:mock} for an image of the galaxy). The morphology of the knots indicate that both B and C have a compact stellar core, while knot A appears as more spread out. This asymmetry in the stellar distribution has led to speculations that the progenitor galaxies are dissimilar (see Ö15). Furthermore, this figure highlights another feature: a string of gas and stars that goes between knot B and A, which then loops back up and to the side. This feature has been dubbed an 'ear' and is reminiscent of a disc arm, which could be interpreted as knot B being the centre of a disc and knot A forming within that disc \citep[see e.g. Ö15;][]{Menacho+19}.

X-ray observations have shown emission from knot B and/or C, which implies the presence of a low-luminosity active galactic nucleus or a strongly emitting X-ray binary system \citep[][]{Hayes+07, Prestwich+15, Gross+21, Danehkar+24}. The possible existence of an active galactic nucleus in knot B would indicate that it was once the bulge of one of the progenitor galaxies. This is because a massive black hole, such as an active galactic nucleus, would not have had sufficient time or mass to form outside the galactic centres during this interaction. Although, it is possible that the black hole was ejected during a previous interaction \citep[see e.g. ][and references therein]{Condon+17}. If knot B is found to contain an AGN, it imposes a morphological constraint on the current positions of the progenitor galaxies' centres. However, this is not conclusive as the black hole could have been ejected during previous interactions.

\citet{LeReste+24} presented observations of the 21 cm \HI line using the {\small MeerKAT} telescope, which uncovered the presence of a tidal tail extending approximately 40 kpc from the East side of the galaxy. The detection of a single tidal tail suggests that one of the galaxies is a disc undergoing prograde motion during the interaction. A prograde orbit has its (disc) rotation aligned with the orbital path of the object it is interacting with. This results in a concentrated region of the disc experiencing the same tidal forces for a longer duration, which yields more pronounced tidal stripping \citep[see Section 3 in][ for details and illustrations]{DucRenaud2013}. In contrast, a retrograde motion, which has the opposite spin direction as prograde, has a more brief close interaction throughout the disc, resulting in a weaker (or no) tidal tail. Additionally, an elliptical galaxy has no axisymmetric rotation (disc) and is unaffected by the stripping from this alignment effect.

Other than the morphological features, Haro 11 exhibits complex kinematics that is common in mergers. The most direct approach for comparison is to compare how different morphological parts of the galaxy move. For example through velocity maps of the inner morphology (e.g. the motion of the gas/dust ear towards us) or the expansion of the tidal tail. Additionally, there are also conical features in the ionised gas around the knots that show clear filaments and outflows \citep[][]{Menacho+21}, suggested to be caused by feedback. As the features of feedback are too detailed and occur spuriously, they can not be directly compared with simulations. However, analysing the average properties of ionised gas and outflows could offer useful comparisons, but this is not within the scope of this project.

\subsection{Recent star formation history}
Haro 11 is currently undergoing a strong starburst with a star formation rate (SFR) upwards of $\lesssim 30\,\Msolyr$ \citep[][]{Hayes+07, Adamo+10, Madden+13, MacHattie+14, Gao+22}, which are derived from different gas tracers and SED fitting methods but with an overall agreement. Recently, \citet[][see also \citealt{Adamo+10}]{Sirressi+22} used spectra to analyse the spectral energy distribution of the three knots in Haro 11 and determined that the energy budget is best modelled by three distinct stellar age populations: $1-4$ Myr, $4-40$ Myr, and $40-100$ Myr. From their analysis they estimate the stellar mass in each knot for the different ages within the range $10^6 - 10^8\Msol$.

One of the striking features of Haro 11 is the leakage of Lyman continuum (LyC) radiation, making it the closest known LyC leaker \citep[][]{Bergvall+06}. The origin of this radiation has predominantly come from knot B and C \citep[][]{Prestwich+15, Gross+21, Ostlin+21, Komarova+24, Danehkar+24}, with \citet[][]{Komarova+24} reporting knot B to have a higher LyC emission and production but lower escape fraction. While radiative transfer is beyond the scope of this study (but will be addressed in Ejdetjärn et al. in prep.), the presence of LyC radiation indicates an absence of neutral gas and dust covering knot B and C. This depletion could be the result of several mechanisms: gas being stripped during the interaction between the progenitor galaxies \citep[][]{LeReste+24}, a displacement of stars relative to the gas during the close encounters (Ejdetjärn et al. in prep.), strong stellar feedback (driven by starburst) forming escape channels \citep[e.g.][]{Menacho+19}, or it may be linked to the intrinsic characteristics of the progenitor galaxies (e.g. dust-poor). Based on the gas distribution around the knots, some authors have proposed that Haro 11's progenitors could have been a dwarf disc galaxy and a less gas-rich galaxy, such as a dwarf spheroidal galaxy \citep[e.g.][]{Adamo+10, Ostlin+15, Menacho+21}.

Extending the analysis beyond the three knots, some observations find notable stellar populations around $\gtrsim 100$ Myr, and as old as 1\,Gyr \citep{Chandar+23, PapaderosOstlin2023}. Relevant for this study, build-up of stellar masses at specific ages in galaxies undergoing a merger might indicate starburst events triggered by close passages during the interactions \citep[e.g.][]{Bournaud+14}. These ages can then serve as constraints for the time between the interactions in our simulated merger.

\subsection{Gas fraction}
The gas mass around the central part of Haro 11 has been estimated for  the molecular $M_{\rm H_2}=2.5\times 10^8-3.8\times 10^9\Msol$\citep[][]{Cormier+14, Gao+22}, neutral \HI $M_{\rm HI}=0.5\times10^9\Msol$ \citep[][]{MacHattie+14, Pardy+16, LeReste+24}, and ionised $M_{\rm H\alpha}= 1.4\times10^9\Msol$ \citep[][]{Menacho+19} gas phases. The stellar mass is around $M_*=1.6\times10^{10}\Msol$ (Ö15), which yields a total gas fraction $f_{\rm g}\sim 14-27\,\%$; depending on the molecular mass. The neutral \HI mass was recently reassessed by \citet{LeReste+24}, who showed that previous observations had not been sensitive enough to detect the tidal tail, which they note makes up $45^{+19}_{-21}$\,\% of the total \HI mass, $M_{\rm HI}=1.11\times10^9\Msol$. The detection and mass-determination of a stellar tidal tail could possibly change the galaxy's gas fraction further. The mass of specific gas phases provide weak constraints to our models, as the formation of e.g. molecular and ionised gas are highly sensitive to the specific feedback and star formation recipes. However, the total gas mass and gas fraction only depend on the star formation physics, which we confirm result in a SFR similar to observations (see Section~\ref{sec:sfr}).

\begin{table}
	\centering  
	\caption{The initial mass and size properties of the two progenitor galaxies. One of the galaxies is initialised with a rotation corresponding to a prograde interaction while the other has a retrograde interaction, and are differentiated by this property. The specifics of how these properties are initialised is described in Section~\ref{sec:simulations}. }  
	\begin{tabular}{l l l l l}
		\hline
		 & & \multicolumn{2}{c}{Galaxies}  \\
		Component & Parameter & Prograde & Retrograde \\
		\hline 
            Halo & Mass $[{\rm 10^9\ M_\odot}]$ & 200 & 167 \\
            & Scale radius [kpc] &  5.12 &  4.74 \\
            & Cut-off radius [kpc] & 60.0  &  55.0  \\
            
		\hline 
            Gas disc & Mass $[{\rm 10^9\ M_\odot}]$ & 5.5 &  1.2   \\
            & Scale length [kpc] &  2.18 & 0.93  \\
            & Cut-off length [kpc] & 7.0  &  4.5 \\
            & Scale height [kpc] &  0.18 & 0.15  \\
            & Cut-off height [kpc] & 0.7 &  0.45  \\
                        
		\hline 
            Stellar disc & Mass $[{\rm 10^9\ M_\odot}]$ & 4.57  & 2.57 \\
            & Scale length [kpc] &  1.2 & 0.9 \\
            & Cut-off length [kpc] &  3.6 & 2.7 \\
            & Scale height [kpc] &  0.24 & 0.1 \\
            & Cut-off height [kpc] &  0.72 & 0.27 \\
                        
		\hline 
            Stellar bulge & Mass $[{\rm 10^9\ M_\odot}]$ & 4.0 & 2.28 \\
            & Scale radius [kpc] & 0.2  &  0.1 \\
            & Cut-off radius [kpc] & 0.35  &  0.13  \\
		\hline
	\end{tabular}
	\label{tab:ics}
\end{table}

 \begin{table*}
	\centering  
	\caption{The orbital parameters and initial inclination of the two galaxies.  }  
	\begin{tabular}{l l l l l l l l l l l}
		\hline
		  & \multicolumn{3}{c}{Position [kpc]} & \multicolumn{3}{c}{Velocities [\kmsecalt]} & \multicolumn{3}{c}{inclination}  \\
		 Galaxy &  $x$ & $y$ & $z$ & $v_x$ &  $v_y$ &  $v_z$ &  $\hat x$ & $\hat y$ & $\hat z$ & \\
		\hline 
            Prograde & +11.40 & -32.77 & +50.42 & -8.94 & -1.81 & -15.72 & +0.7138 & -0.6680 & -0.2105 \\
            Retrograde & -16.22 & +32.88 & -50.05 & +8.94 & +1.81 & +15.72 & -0.6710 & +0.2138 & +0.2048 \\
		\hline
		\hline
	\end{tabular}
	\label{tab:ics_orbital}
\end{table*}

\section{Numerical simulations and methods} \label{sec:simulations}
In this section we outline the physical recipes for star formation and feedback, and describe our method for determining the initial properties and orbital parameters of the progenitor galaxies. The simulations presented in this work were performed with the hydrodynamics and $N$-body code \ramses \citep[][]{Teyssier2002}, which uses an Adaptive Mesh Refinement (AMR) grid method with an HLLC Riemann solver \citep[][]{ToroSpruceSpeares94}. The code solves the conservative Euler equations for the dynamics of the gaseous component using a second order Godunov scheme. We adopt an ideal mono-atomic gas with adiabatic index $\gamma = 5/3$. The Euler equations consider the hydrodynamics and the gravitational effects from stars, dark matter, and fluid self-gravity. Dark matter and stars only interact through gravity, as collisionless particles. We employ a sub-grid recipe for the formation of stars and their feedback, which we will describe briefly in the coming section and refer to \citet{Agertz+13, Agertz21} for details.

The aim of this study is to explore possible formation scenarios for the Haro 11 galaxy. To this end, we have performed roughly 500 low-resolution simulation tests of a merger between two disc galaxies, as part of an iterative trial-and-error parameter study to approach the orbital and galaxy parameters that agree with Haro 11 observations. To reduce the run time, no stellar feedback was included in the initial test simulations. For each simulation we varied various parameters relevant for the progenitor galaxy orbits (e.g. velocity) or their intrinsic properties (e.g. disc size) to make the simulation provide a qualitative visual match to the observed morphology, SFR, kinematics, and stellar properties observed in Haro 11. Parameter changes that improved the model were kept and further iterated upon. However, there is no definitive best set of parameters as orbital and galactic parameters interact complexly, causing parametric degeneracies where significantly different sets of initial conditions can achieve similar results.

\subsection{Star formation and feedback physics}
In our simulations, the formation of stars is treated as stochastic events, with the amount of stars formed sampled from a discrete Poisson distribution and following the Schmidt law \citep[][]{Schmidt59, Kennicutt98}
\begin{align}
    \dot\rho = \epsilon_{\rm ff} \frac{\rho_{\rm g}}{t_{\rm ff}}\ {\rm for}\ \rho_{\rm g} > \rho_*,
\end{align}
where $\rho_*=100\,{\rm cm}^{-3}$ is the density threshold for star formation, $\rho_{\rm g}$ is the gas density, $t_{\rm ff} = \sqrt{3\pi/32 G \rho}$ is the free-fall time for a spherical cloud, and $\epsilon_{\rm ff}$ is the star formation efficiency per free-fall time. Star particles follow a \citet{Chabrier03} initial mass function with a total mass $10^3\Msol$. We adopt $\epsilon_{\rm ff} = 10\%$, as this choice in galaxy simulations has been shown to reproduce the same low efficiency ($\epsilon_{\rm ff} \sim 1\%$) observed in giant molecular clouds \citep[see e.g.][and references therein]{AgertzKravtsov16, Grisdale2019}.

Stellar feedback follows the subgrid recipe detailed in \citet[][see also \citealt{AgertzKravtsov15}]{Agertz+13} and tracks the injection of momentum, energy, and metals of supernovae Type Ia, Type II, and stellar winds. Individual supernovae are considered resolved when the host cell is resolved by at least six cooling lengths, following \citet{KimOstriker15}, and is otherwise deposited in the neighbouring gas cells. Metallicity is traced through oxygen and iron, which are combined into a total metallicity according to the relative solar abundance \citep[][]{Asplund2009}.

\subsection{Fiducial progenitors setup}\label{sec:ICs}
We based our initial guess of the galaxies' orbital parameters and inclination on the initial conditions of a simulation of the Antennae galaxies by \citet{Renaud+15}. Ö15 made a comprehensive comparison between Haro 11 and the Antennae galaxies and found similarities between their morphology, kinematics, and stellar properties. Both galaxies are currently undergoing a merger, but with the Antennae galaxies being larger and more massive. The simulations by \citeauthor{Renaud+15} manage to successfully recreate the morphology and several important properties of the Antennae. However, these galaxies have several observed differences between them and the initial conditions need to be tweaked to account for these.

Firstly, recent \HI observations indicate that Haro 11 only has one visible tidal tail \citep[][]{LeReste+24}, compared to two in the Antennae galaxies. The strength of a tidal tail can be weakened by flipping the rotational axis (inclination) of one of the progenitor galaxies to form a retrograde orbital motion, but this does not guarantee that the tidal tail will be removed, as other properties interact in a complex way to affect its strength and shape. Notably, the Antennae galaxies are a symmetric interaction (in terms of mass, size, and morphology), while Haro 11 has been hypothesised to be the result of dissimilar progenitor galaxies (see the discussion in Section~\ref{sec:obs_constraint}). For the initial setup we reduced the gas mass and size of the retrograde galaxy (as is detailed below), which is sufficient to reduce the strength of the tidal tail to be completely undetectable (see Section~\ref{sec:match_tidal_tail} for details on this process).

The Haro 11 galaxy is roughly one third the size of the Antennae galaxies, and we adopt this size scaling between our simulations and the Antennae simulation by \citet{Renaud+15}. In their comparison, Ö15 indicate that the galaxy might be 4 times as dense as the antennae, which is apparent from the similar gas mass, but smaller size, between our prograde galaxy and NGC 4038 in the Antennae simulation. In Section~\ref{sec:match_tidal_tail}, we show an example of how varying the size and density of the progenitor discs affects the final morphology.

\subsubsection{Initial conditions}\label{sec:modelling_disc}
Both progenitor galaxies were modelled as disc galaxies with a NFW \citep[][]{NavarroFrenkWhite96} halo profile for the dark matter, a spherical \citet{Hernquist1990} profile for the stellar bulge, and exponential profiles for the stellar and gas discs. The initial position and velocity of each particle and gas element was then generated by {\small MAGI} \citep[][]{MikiUmemura2018} using these profiles to distribute $10^6$ dark matter particles, $10^6$ stellar particles in the disc, and $10^5$ particles for the stellar bulge. The gas disc was initialised as $10^6$ particles, which were then deposited onto the AMR structure.

Table~\ref{tab:ics} lists the properties of each component in both progenitor galaxies, while Table~\ref{tab:ics_orbital} provides their orbital parameters and inclinations. The two progenitor galaxies are labelled \emph{prograde} and \emph{retrograde}, to differentiate based on which galaxy forms the tidal tail. The initial disc rotation velocities were calculated from each galaxy's total mass profile and oriented according to their inclinations. The galaxies were positioned sufficiently distant to prevent any overlap between their respective halos, allowing a period of secular evolution before any significant tidal interaction.

In order to determine the initial mass budget, we assumed the galaxies have retained a mass baryon fraction $f_{\rm b}$ close to the cosmic baryon fraction $\Omega_{\rm b}$, i.e. $f_{\rm b} = (M_{\rm g} + M_*)/(M_{\rm g} + M_* + M_{\rm DM}) \approx 15.6\%$ \citep[][]{Planck+20}. We base the initial stellar mass on observational constraints and do abundance matching \citep[following Figure 9 in][]{Behroozi+19} to determine the mass of the dark matter halo. The total stellar mass of Haro 11 is on the order of a few $10^{10}\Msol$ (\citealt{Ostlin+01}; \citealt{Madden+13}; Ö15). The resulting dark matter matches with the total dynamical mass $\sim 10^{11}\Msol$ of Haro 11, estimated by Ö15 through dynamical reasoning of velocity field observations in the outer part of the galaxy.

The initial gas mass of the galaxies was calculated as the total observed mass between the molecular, neutral, and ionised gas phase (see Section~\ref{sec:obs_constraint} for values). We then added $1.25\times10^9\Msol$ to this mass, in order to offset the gas consumed through star formation during the simulation run (determined during test runs). Similarly, we removed the same amount of mass from the observed stellar mass to determine our initial stellar mass. Additionally, some of this gas mass budget is added to the circumgalactic medium (see below). The gas mass in the discs is added in the simulation as neutral gas, and any additional gas phases naturally form as the galaxy evolves. The initial metallicity of the gas discs was directly taken from observations ($12+{\rm [O/H]}=7.9$; Ö15) and translated to a total metallicity $Z=0.2\,Z_\odot$.

The simulation was started at a low spatial resolution to allow the initial conditions to relax and then increased after a few hundred Myr, with plenty of time before the first interaction. The maximum spatial resolution of the fiducial simulation is $\Delta x = 36$\,pc, and is within a 300$^3$ kpc$^3$ volume. The circumgalactic medium spans the entire simulation volume with an initial temperature $T\sim10^8$\,K and hydrogen number density $n_{\rm H} = 10^{-4}\,{\rm cm}^{-3}$, which is within the densities observed of local galaxies and in simulations \citep[][]{Machado+18}. A proper gas halo is necessary to produce realistic outflows and tidal tails, and to ensure the gas mass of the system matches the cosmic baryon fraction for the given stellar mass. However, the circumgalactic gas content varies widely between systems and is not well constrained. We constructed tests with higher and lower circumgalactic densities, which is briefly discussed in Section~\ref{sec:cgm_test}.

\subsection{Strategy for tweaking initial conditions}\label{sec:strategy_match_obs}
This section details how the initial conditions for the two progenitor galaxies were improved upon through iteration. While our approach is similar to a parameter study, the focus of this project is to compare the fiducial Haro 11 model with observational data and discuss possible merger scenarios for the galaxy and its properties. We will not go into detail on the full range of parameters and their agreement with each observed property. Instead, we will here briefly discuss our approach for narrowing down the range of the parameter space, and in Section~\ref{sec:extra_tests} we show visual examples of some early low-resolution tests. These tests demonstrate how some specific changes to the initial conditions affect the morphology of the resulting galaxy. 

The main observed properties used for constraining our model were: gas and stellar morphology, SFR, gas kinematics, and stellar masses and ages. To match these, we varied for each progenitor galaxy these properties: positions, the velocities (speed and angle), gas/stellar mass content, inclination, and size of the individual galaxies. In particular, we explored the relative mass ratio and gas fraction between the galaxies, as these parameters are connected to the formation of tidal tails, and the gas fraction has been suggested to be different for the progenitors (see discussion in Section~\ref{sec:obs_constraint}).

The orbit between the two galaxies is strongly entangled with all of the properties we want to compare with observations. Adjusting the velocity is the main control we have of the orbit, including the timing and proximity of the close interactions. The initial speeds were altered between half and twice the initial speed of the fiducial presented in Table~\ref{tab:ics_orbital}. When evaluating the velocity angle we (primarily) changed the $x$ and $y$ velocity components, and with the aim to fine-tune the tidal tail and inner morphology.

\begin{figure*}
    \includegraphics[width=0.95\textwidth]{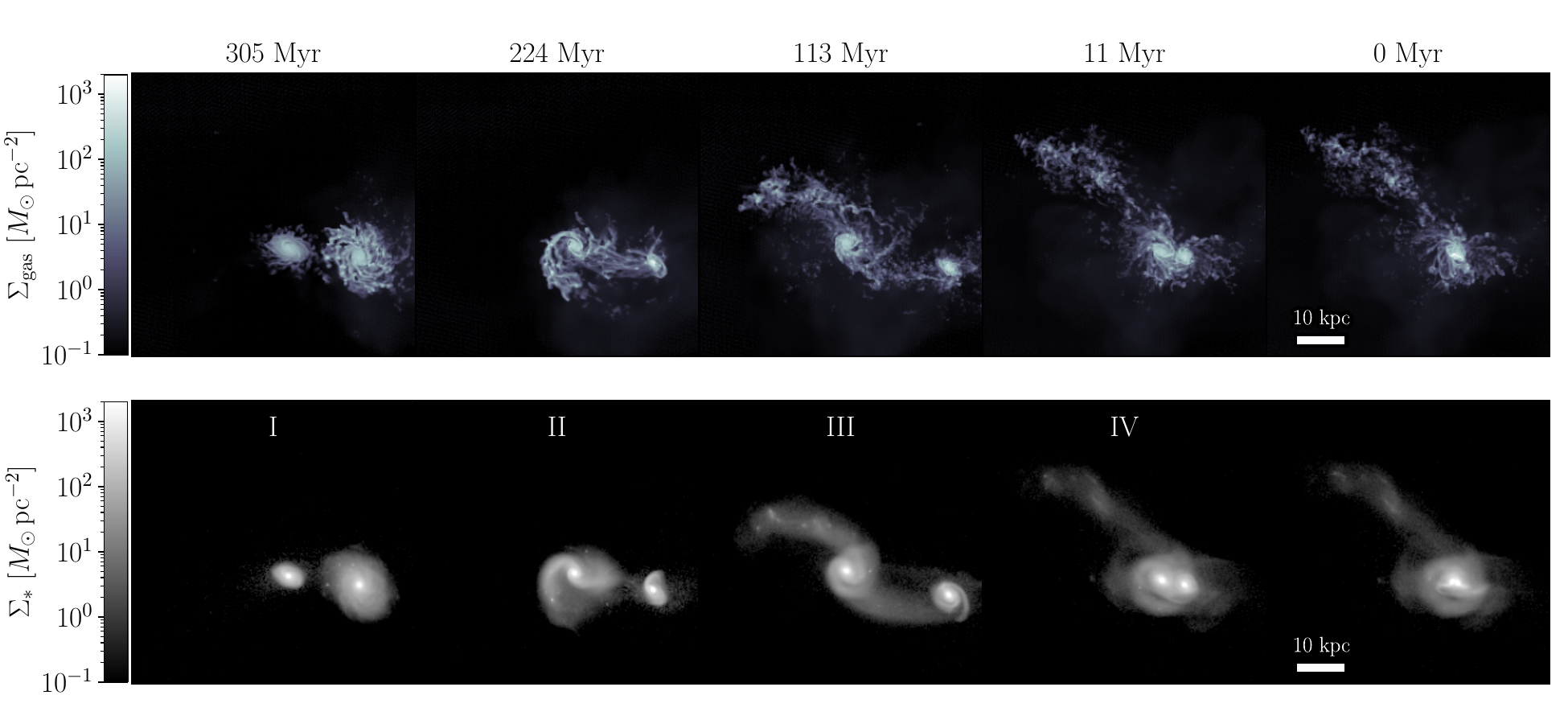} 
    \caption{ The column density of gas (top row) and stars (bottom row) for the large-scale structure of the galaxy are shown at various time outputs, spanning from the initial interaction to the point where the galaxy most closely resembles Haro 11. The plots marked with Roman numerals correspond to the times in the SFR plot in Figure~\ref{fig:sfr}. }
    \label{fig:timeline}
\end{figure*}

Other than the velocity, the mass ratio between the galaxies has a direct impact on the orbit. However, compared to the velocity, the masses are not as easy to fine-tune and have additional effects on other observables. For example, the SFR depends on the orbit but the amount of stars formed in the individual galaxies depends on their halo properties in a non-linear way. Thus, the mass ratio was primarily changed in early tests.

The tidal tails were formed during the first encounter between the progenitors, which is closely connected to their initial orbit. Additionally, the inclination of each disc has a significant impact on the strength and morphology of the tidal tail (see discussion of prograde vs retrograde in Section~\ref{sec:obs_constraint}). Further examples of how these parameters affect the appearance of the morphology are provided in Section~\ref{sec:match_tidal_tail}. Additionally, the inner morphology of the galaxy is also dependent on these parameters, as well as the initial mass and size of the discs.

Our strategy at each iteration was as follows. Most tweaks were done one feature at a time. We first improved upon the match with the large-scale and general properties, i.e. tidal tail properties (morphology, kinematics, stripped mass), stellar populations, and SFR, before focusing on the inner region. This was done by tweaking the initial velocities, inclination, and relative mass ratio. For the inner part of the merger, we fine-tuned the stellar and gaseous morphology of the knots and the "ear" by \emph{slightly} shifting the velocity, inclination, and galactic properties; this of course has a small impact on the tidal tail properties, but these were minor relative to the improved match in the inner region.

As observations of the tidal tail position already aligned well with the $z$-axis of the fiducial setup, the viewing angle only required a relatively small adjustment from this axis, and was done last. The choice of viewing angle was done to position the knots at the expected positions in the projected 2D space (relative to each other and the tidal tail), while also keeping the tail position at a good match with observations.

\subsection{Analysing the simulation output}\label{sec:analysis}
The total runtime of the fiducial simulation is 1280 Myr, which is just after the centres of the two galaxies fully coalesce. We analyse the galaxy at the time output we deem to be closest in terms of SFR and morphology to the Haro 11 galaxy. We define the snapshot that best matches the observations (i.e. current day) as 0 Myr, during the second close interaction. The morphology and kinematics we calculate are largely invariant on the choice of time by several Myr, but the SFR can vary by a factor of 2-4 within only a few Myr. After $\gtrsim10$\,Myr, the morphology and SFR are notably different to Haro 11. 

The exact morphology and kinematics of the inner galaxy is dependent on the viewing angle we adopt. Our approach is to find an angle that improves the match for the inner morphology (i.e. relative position between knots and the 'ear') while keeping the tidal tail at roughly the correct relative angle to the inner morphology. The fiducial viewing angle is in-between the $z$ and $y$ axis of the simulation (specifically, $\hat{n}=[0.26, -0.64,  0.72]$). The morphology is insensitive to a change in angles within a subtended angle $\sim10^\circ\times10^\circ$, with some directions being less or more affected. 

The SFR is calculated by binning stars by stellar age, summing the stellar mass within each time bin, then subtracting the stellar mass of a neighbouring bin and dividing by the bin size. Our results are presented with a 1 Myr bin width to highlight the quick variations during the interactions. However, if adopting 10 Myr bins, which is equivalent to the timescale for \Halpha observations, the overall SFR history was found to be very similar.

\begin{figure*}
    \includegraphics[width=0.40\textwidth]{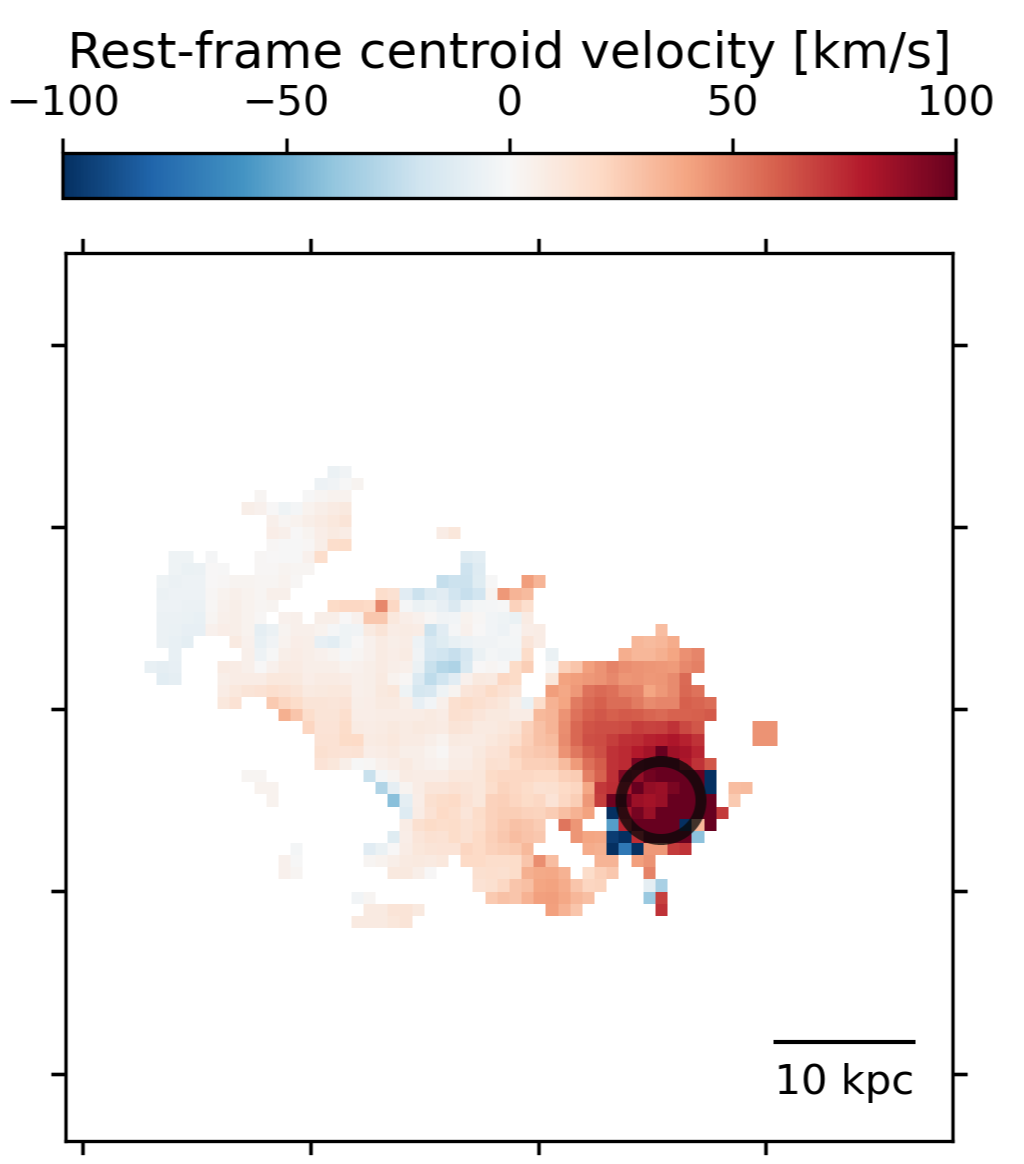}
    \includegraphics[width=0.45\textwidth]{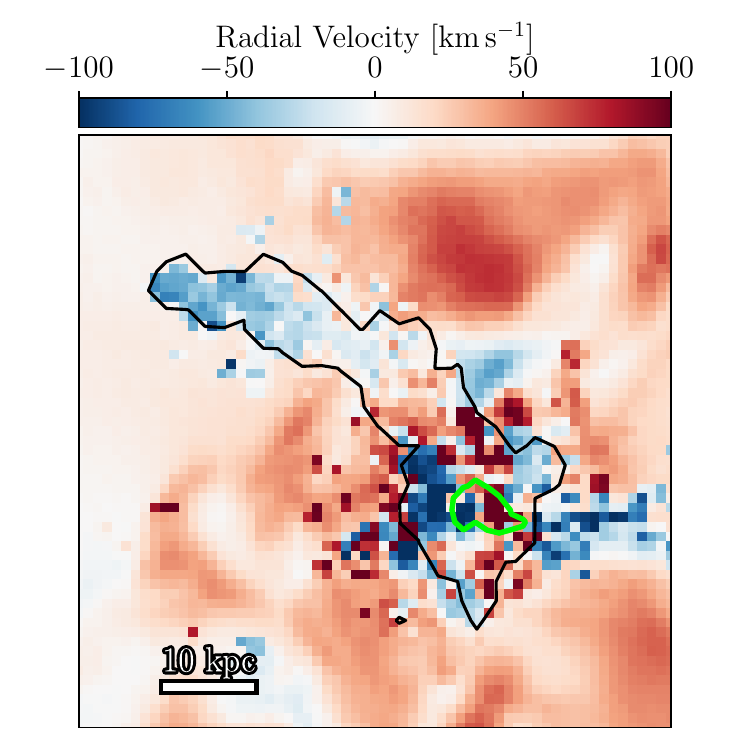} 
    \caption{ Velocity maps of the \HI tidal tail. Left panel shows the 21-cm line, in emission only, of Haro 11 with {\small MeerKAT} \citep[][see their paper for specifics of the observation and data reduction]{LeReste+24}. Right panel shows the simulation; the green contour outlines the central stellar component, and the black contour outlines the gas column density exceeds 10\% of its maximum value.   } 
    \label{fig:tidal_tail_velocity}
\end{figure*}

\begin{figure*}
    \includegraphics[width=0.48\textwidth]{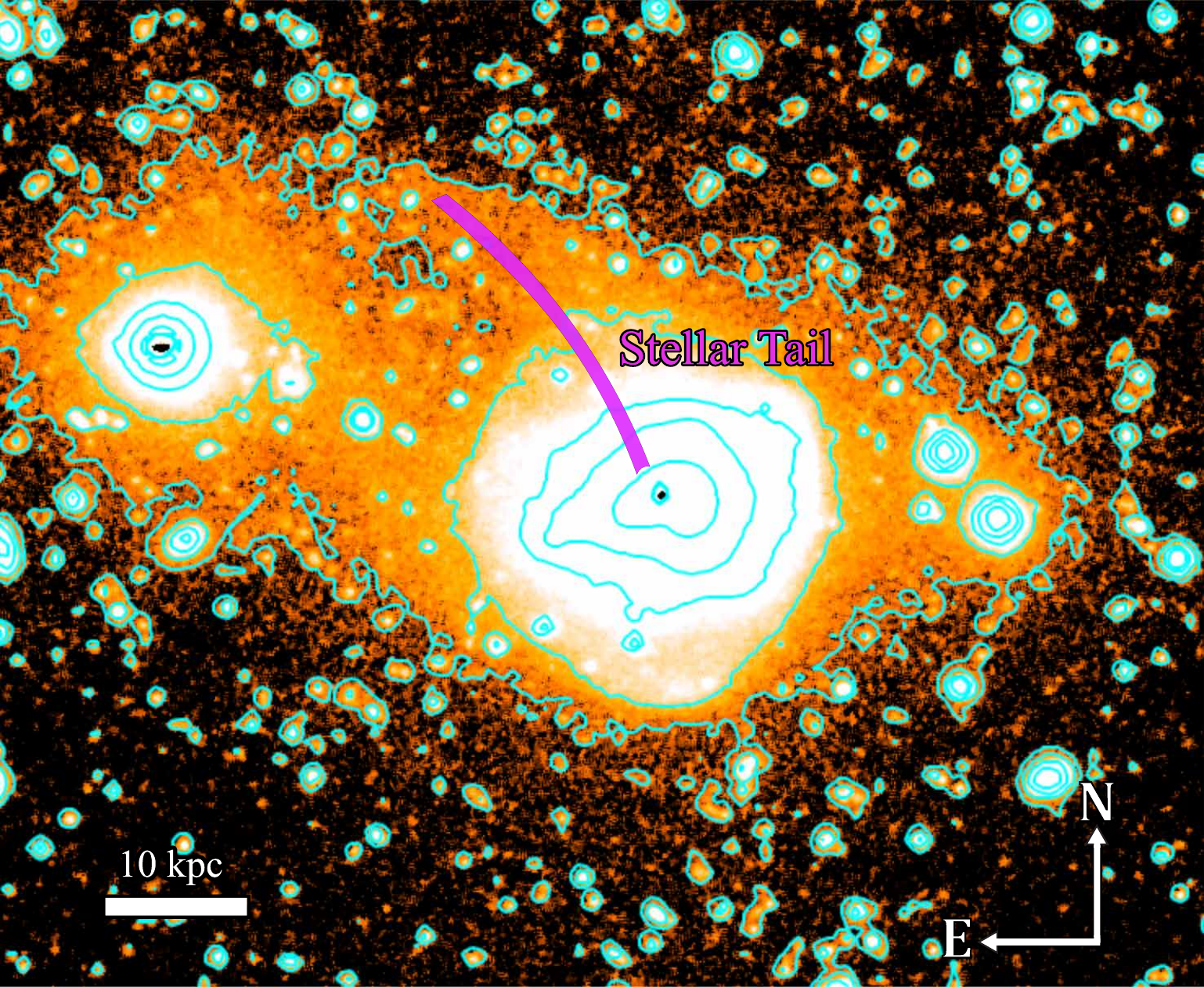}  
    \includegraphics[width=0.48\textwidth]{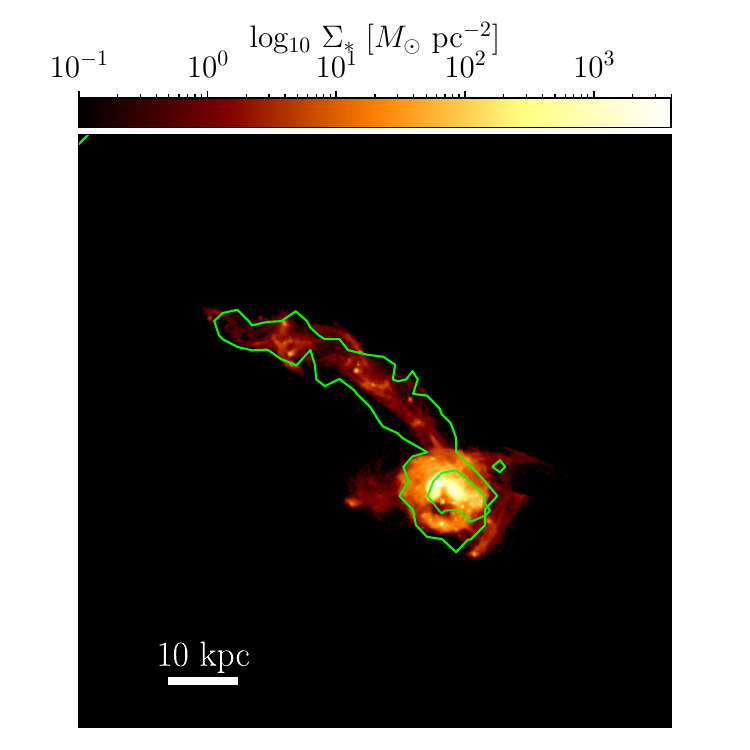} 
    \caption{ The stellar tidal tail from observations (\emph{left}) and simulations (\emph{right}). The observations are in the B-band with VLT/FORS (see details in Section~\ref{sec:tidal_tail}), with cyan contours showing the stellar distribution for different thresholds.  The simulation has green contours that highlight the gas distribution of the tidal tail and the inner region. } 
    \label{fig:stellar_tidal_tail}
\end{figure*}

\begin{figure*}
    \includegraphics[width=0.98\textwidth]{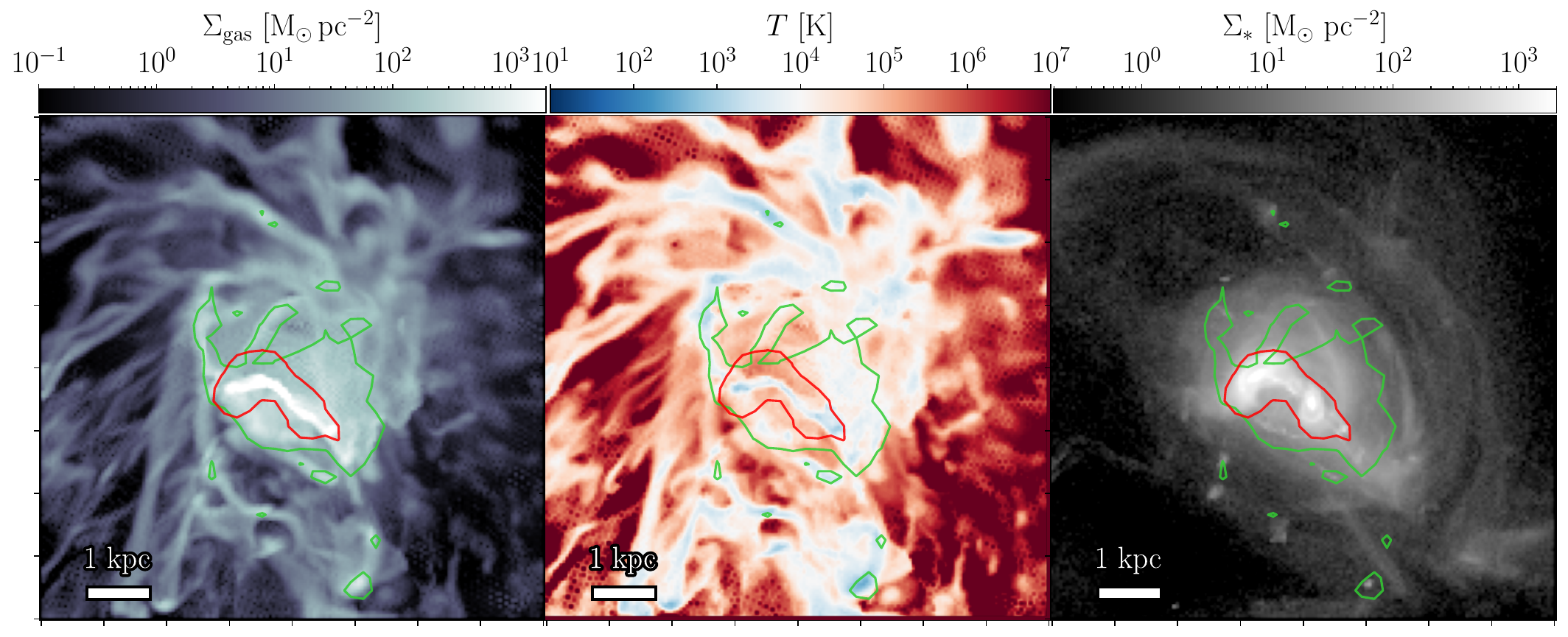}
    \caption{Projections maps of the inner part of the galaxy at the output where the galaxy most closely resembles Haro 11. The column densities are calculated as the sum of mass within a pixel and then divided by the pixel area. The temperature is the mass-weighted mean. Green/red contours correspond to the most dense regions of gas/stars. } \label{fig:dens_temp_stell}
\end{figure*}

\begin{figure}
    \centering\includegraphics[width=0.38\textwidth]{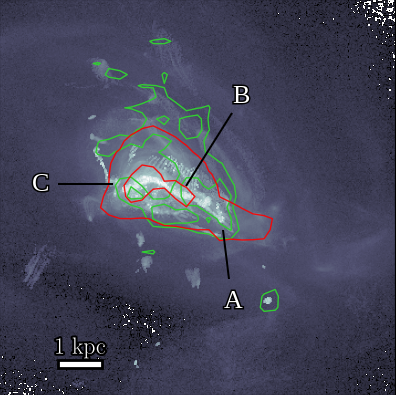}
    \caption{ Projection map of the stellar mass-weighted by their inverse stellar age, $M_{\rm *,\,young}=M_* \times (t/4\,{\rm Myr})^{-1.25}$, to represent young, bright stars. The red/green contours correspond to the projected stellar/gas column density. Combined with Figure~\ref{fig:mock}, in which knot A is not readily observed due to this extinction, this map demonstrates that bright stars form in knot A but are obscured by dust. }
    \label{fig:stellar_weight_func}
\end{figure}

\section{The fiducial Haro 11 model}\label{sec:haro11_model}
In this section we present our fiducial Haro 11 model, chosen from among a large set of simulations (following the method in Section~\ref{sec:ICs}) as one of the best matches by eye with various key features of Haro 11. In particular, we compare the morphology (inner and tidal tail), SFR, stellar properties, and kinematics in our simulation with observational data. This simulation setup offers a broad agreement between all these different properties, which indicates it serves as a good basis to describe the formation scenario of Haro 11. We discuss the strengths of our fiducial model as the origin of Haro 11 and apply it to interpret observations in Section~\ref{sec:discussion}, along with discussing possible improvements for future numerical work. 

Additionally, we present the current stellar and gas masses, as well as the SFR, of this simulation in Table~\ref{tab:galaxy_properties}. The gas mass is also divided into phases by weighting the mass by phase-specific variables, such as \Halpha emissivity \citep[see][for details]{Ejdetjarn+22}: molecular, neutral, and ionised, in order to compare its distribution with observations. However, we note that the gas phase masses are uncertain, as they are sensitive to the stellar formation and feedback recipes adopted. Additionally, ($\sim 1.9\times10^9\Msol$) of hot/warm ionised gas within the halo of our galaxies, which is too diffuse to be observed and therefore not added to the ionised gas mass budget.

 \begin{table}
    \centering  
    \caption{ Properties of the galaxy from the fiducial Haro 11 simulation, analysed at current time, when the morphology and SFR best match observations, for a volume of 60\,x\,60\,x\,60\,kpc$^3$ around the central part of the galaxy. We separate between ionised gas around the galaxy ($<10$\,kpc) and ionised CGM gas, as the latter has no observational comparison. The CGM is, thus, excluded from the gas fraction and total gas mass. }  
	\begin{tabular}{l l l l}
    	\hline
    	Stellar mass [$10^9\Msol$]          &  16.49 \\ %
            Total gas mass [$10^9\Msol$]        &  4.33  \\ %
            Gas fraction [\%]                   &  20    \\
            Molecular gas mass [$10^9\Msol$]    &  1.35  \\ 
            Neutral gas mass [$10^9\Msol$]      &  1.06  \\ 
            Ionised gas mass [$10^9\Msol$]      &  1.77  \\ 
            Ionised CGM gas mass [$10^9\Msol$]      &  1.90  \\ %
            SFR [$\Msolyr$]                     &  43 \\ 
    	\hline
	\end{tabular}
	\label{tab:galaxy_properties}
\end{table}

\subsection{Tidal tail}\label{sec:tidal_tail}
The first step of analysing the simulation is to visualise the merger process and the tidal tail, as it is one of the strongest signs of Haro 11 being a merger. In Figure~\ref{fig:timeline} we present several projection maps of the gas and stars as part of a timeline of the simulation. These maps follow the formation of the tidal tail, from the first close interaction until they reach the current observed time of Haro 11, which is $\sim10$ Myr into the second interaction; a time period of roughly $\sim 200$\,Myr. Furthermore, the maps have Roman numerals annotated, corresponding to time markers in the figure of SFR as a function of time in Section~\ref{sec:sfr}, which highlights the burst of star formation during the close interactions. 

This simulation matches the length and kinematics of the \HI tail presented in \citet{LeReste+24} using {\small MeerKAT} observations of the 21-cm line\footnote{The reduced MeerKAT 21\,cm data is publicly available at: https://archive-gw-1.kat.ac.za/public/repository/10.48479/7zn6-bw59/index.html}. The tidal tail is extending in the same direction relative to the inner morphology as in Haro 11 (north-east) and has the same length $\sim 40$\,kpc. Furthermore, we find that the relative \HI mass inside the tidal tail is $\sim 36\,\%$, which is within the observed value $44^{+20}_{-19}$\,\%.

A velocity field map of the \HI observed in emission \emph{only} is presented in the left panel of Figure~\ref{fig:tidal_tail_velocity} (the equivalent plot for absorption can be found in Appendix~\ref{app:tidal_tail_abs_map}), and the \HI velocity in our simulation is on the right. Most of the \HI emission is red-shifted relative to the central galaxy, for which \citet{LeReste+24} adopt the optical redshift from \citet[][]{Bergvall+00}; with similar systemic velocity as the ionised gas around the knots \citep[see][]{Menacho+21}. Observations indicate the tidal tail is very slightly blue-shifted at a maximum speed of $\lesssim -20$\kmsecalt, while our simulation has a tidal tail with $\lesssim -50$\kmsecalt. This discrepancy is most directly explained by how the zero-velocity is set in the simulations, as there is no clear systemic velocity in a galaxy merger. Other possible factors are the viewing angle of the simulation or that the edges of the tidal tail are too faint to be detected by observations. We discuss this in more detail in Section~~ \ref{sec:discussion}. 




We next present new optical observations with VLT/FORS, obtained under program 113.26LV, designed to specifically investigate the presence or not of a faint tidal tail in Haro 11. The galaxy was observed for 2 hours each in the B (\texttt{b\_High}) and I (\texttt{I\_Bessel}) filters. These filters were chosen since they do not contain bright emission lines (at the redshift of Haro 11, H$\beta$ and the bright [OII] and [OIII] lines are not transmitted) and are hence primarily probing stellar emission. We here focus on the B-band image since due to the dark sky it better reveals the outer low surface brightness structures of Haro 11. Details about the reduction and calibration is given in Appendix~\ref{app:stellar_tidal_tail}. In the left panel of Figure~\ref{fig:stellar_tidal_tail} we show a cutout from the final B-band image, where we detect a tidal extension from the north (N) part of Haro 11 towards E-NE (the angle measured from the N-axis is $\sim65$ deg), approximately corresponding to the orientation of the \HI tail. The outer tip of tail identified in B-band extends to $\sim 30$ kpc from the centre. In addition we see a tidal feature from the upper part of the galaxy, extending towards W-SW, and another feature extending from the southern part towards E-NE (in the direction towards a bright foreground star). The latter two features are somewhat affected by the extended point spread functions from nearby bright stars, making their total extent uncertain, but are real. The features are present also in the FORS I-band image, but at lower significance. In addition, an archival HST/WFC3/IR/F160W image confirms the existence of these features at low S/N but the field of view (2.7" x 2.7") is too small to well cover the Haro 11 tidal system. Taken together, we now have firm evidence that Haro 11 also possesses low surface brightness stellar tidal tails. The E-NE tail has a surface brightness of $\sim27.5$ mag/arcsec$^2$ at its midpoint, corresponding to $\sim 7 \times 10^5\,L_\odot\,$kpc$^{-2}$.

The stellar tidal tail is slightly shorter than the \HI component ($\sim$30\,kpc compared to $\sim$40\,kpc). This is likely due to a difference in scale radius or truncation between the two components, i.e. the gas disc having a larger scale radius. We briefly explored the effects of truncation on the disc in Section~\ref{sec:match_tidal_tail}. Additionally, the image of the stellar tidal tail shows clear shell structures around the galaxy. The simulation shows signs of these shells as well, which is encouraging for our model but is not directly applicable as a modelling constraint.

\subsection{Inner morphology}\label{sec:visualise_morphology}
In order to put our simulations into context, we first present an observation of the Haro 11 galaxy on the left in Figure~\ref{fig:mock}. This was presented in \citet{Adamo+10} using HST in three wavebands filters; in particular they use F220W and F814W with the ACS/HRC instrument, and F435W with ACS/WFC. Additionally, the right panel of the figure shows a mock observation of our simulation using the post-processing radiative transfer code \rascas \citep[][]{Michel-Dansac+20} to obtain the stellar continuum and apply the same HST filter profiles as the observations. Dust absorption follows the SMC law \citep[e.g.][]{Laursen+09} and assumes a dust-to-metal conversion factor, which increases with metallicity and cut-off at $T>10^4$\,K. The key morphological features of Haro 11 are annotated: three distinct stellar knots (A, B, C) and a dusty arm going upwards from knot A to the right of knot B (the 'ear'). All of these morphological features can be identified in the mock observation as well, but knot A is not as readily apparent, which is discussed below.

The inner gas and stellar morphology, as well as temperature, of the galaxy at the current observed time is presented as projection maps in Figure~\ref{fig:dens_temp_stell}. The high density gas around knot B and A correspond to areas of high stellar density. A disc arm with gas and stars is seen on the right side of the galaxy in each panel and is reminiscent of the ear observed in Haro 11. Compared to observations, the positions of these features are slightly mismatched in position relative to each other, but are broadly in the expected positions. Additionally, the dense gas that is overlapping with the stellar core on the left (knot C) is behind the core. 

The main reason that knot A is not observed in our mock observations is due to the gaseous and dusty arm covering the stellar knot, but we can infer its presence from the stellar mass and young population. In Figure~\ref{fig:stellar_weight_func} we show a stellar map weighted by the inverse stellar age, as $M_{\rm *,\,young}=M_* \times (t/4\,{\rm Myr})^{-1.25}$, which is an approximate bolometric luminosity weighting; this highlights the presence of young, luminous stars. This figure shows that there is significant mass of young stars around knot A, which would be visible in the absence of extinction. The reason why knot A is more obscured, and other reason why knot A is abscent in our mock image, is discussed in more detail in Section~\ref{sec:discussion}. Additionally, in Figure~\ref{fig:stellar_ages} we present a projection map of the median stellar ages, from which it is also evident that all of the knots, including A, contain many young stars. From this, we infer that knot A likely reside within the disc arm (which also makes up the ear) of the prograde progenitor galaxy, which has its centre at knot B. The knots' properties are presented in more detail in Section~\ref{sec:knot_properties}. %

\subsection{Star formation rate}\label{sec:sfr}
\begin{figure}
    \includegraphics[width=0.47\textwidth]{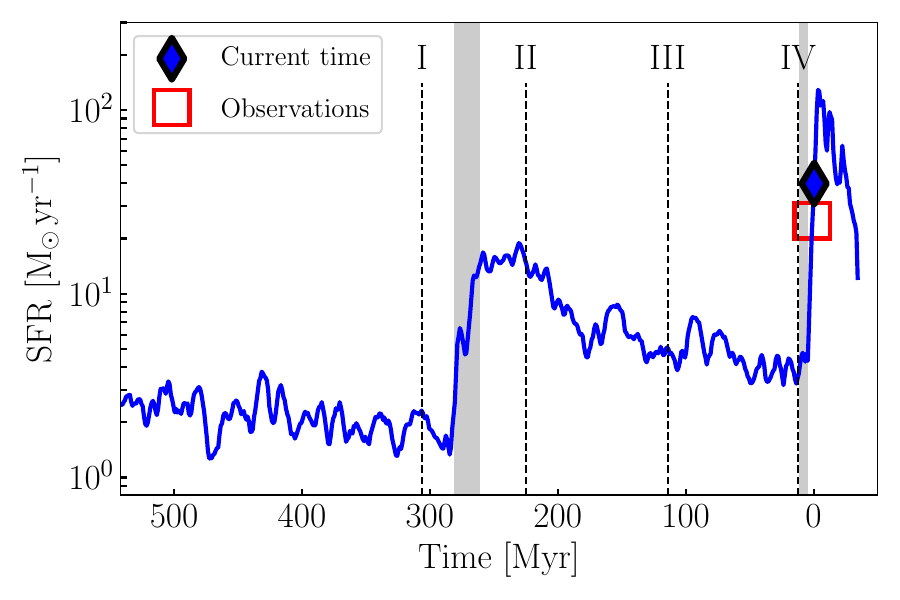} 
    \caption{The star formation rate of the simulation. The time axis is set that the current observed time is at 0 Myr. Only the last $\sim 500$\,Myr are shown, as there is no interaction between galaxies before this and the galaxies have reached a roughly constant SFR (see Section~\ref{sec:analysis} for details). Specific time periods are marked with Roman numerals, corresponding to the projection maps with same number in Figure~\ref{fig:timeline}. The grey shaded areas are the estimated pericentre periods of interaction. The starbursts at 0 Myr and 250 Myr correspond to close, gravitational interactions. The diamond marks the time at which the galaxy best resembles Haro 11, following the criteria in Section~\ref{sec:ICs}, and the square represents the SFR range mentioned in Section~\ref{sec:obs_constraint}. The galaxy is in the middle of its second starburst event, with a ${\rm SFR} = 43\Msolyr$. } 
    \label{fig:sfr}
\end{figure}

\begin{figure}
    \centering\includegraphics[width=0.50\textwidth]{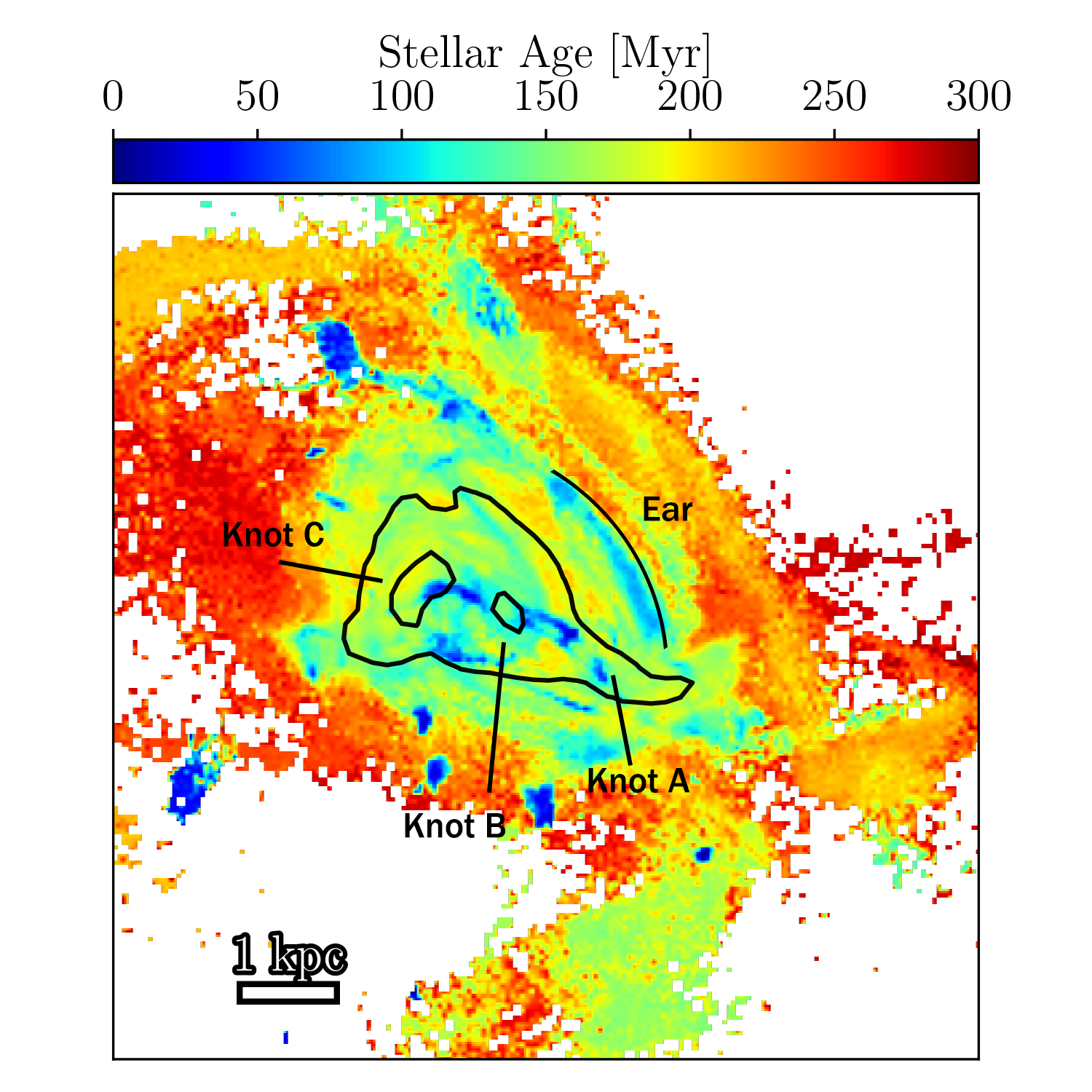} 
    \caption{Projection map of the median stellar ages with black contours of the stellar column density. Labels and lines mark the various features described in Figure~\ref{fig:mock}. }
    \label{fig:stellar_ages}
\end{figure}

Observations show that Haro 11 is currently experiencing a burst of star formation, which is attributed to an undergoing merger. We present the SFR as a function of time in Figure~\ref{fig:sfr}, annotated with Roman numerals as timestamps to the timeline images in Figure~\ref{fig:timeline} and with shaded areas for the time of pericentre passages. The figure shows two peaks in the SFR that are associated with close interaction between the two galaxies. The SFR burst is thus likely caused by interaction effects, e.g. gas compression due to tidal forces, cloud-on-cloud collision, tidal torques that drive gas towards the centres of the galaxies \citep[e.g.][]{Renaud+14, Renaud+19}.

The current day SFR of our fiducial simulation match with the observed values, ${\rm SFR}\lesssim 30\Msolyr$ (see Section~\ref{sec:obs_constraint} for references). Furthermore, the starburst periods in our simulations can be readily compared to observations of stellar populations with distinct ages, which helps constrain the star formation history. The peaks indicate that these stellar populations should have ages of roughly $200-250$\,Myr, from the first interaction, and a few Myr, from the current interaction. We will explore this further in the next section by comparing stellar analysis models of the populations in the three knots.

\subsection{Knot properties}\label{sec:knot_properties}
We next evaluate the age and mass of the stellar knots. The knots are identified as shown in Figure~\ref{fig:mock}, around a 0.5 kpc radius (1 kpc for knot A as it is less concentrated than the other knots). We want to compare with the stellar masses and ages of individual knots attained by \citet{Sirressi+22} from their SED fitting. To this end, we divide the clusters into three populations of age ranges: 1-4 Myr, 4-40 Myr, and 40-300 Myr, which is comparable to the populations in their best fit (we extend the range to 300 Myr to account for uncertainties in the precise time of the first interaction). For clarification, we will not present any detailed quantitative comparison of the stellar masses inside the knots, as the observational uncertainties reported allow a wide range of ages and masses, and the exact mass in our simulations depend on, e.g., the specifics of the model and observed time.

We begin by noting that knot B and C contain a significant amount of old $\gtrsim 100$\,Myr stellar mass, on the order of $\sim 10^8\Msol$, while in knot A this mass is an order of magnitude lower $\sim 10^7\Msol$. Also, the average age of stars in knot A and B are much younger than those in knot C. All these features are in agreement with properties of the three knots reported by \citet[][]{Sirressi+22}. Furthermore, Figure~\ref{fig:stellar_ages} details the median stellar age distribution for the knots, and outlines the various features in the galaxy along with a contour of the stellar mass distribution. Knot B and C are a mix of old and young stars and have a median age of $\sim150$\,Myr, while knot A is primarily young stars. Additionally, the regions between the knots are forming new young stars.

We briefly explore the cluster mass and age distribution in the knots. To identify clustering within our simulations, we employ the friends-of-friends algorithm HOP \citep[][]{EisensteinHut1998} on a 12\,kpc box around the galaxy centre (roughly corresponding to stars observed in e.g. Figure~\ref{fig:dens_temp_stell}). A cluster is defined as having stellar peak densities $50\Msol\,{\rm pc}^{-3}$, a minimum density to merge groups as $50\Msol\,{\rm pc}^{-3}$, and outer density threshold of $0.1\Msol\,{\rm pc}^{-3}$; set at a low value to detect the spread out clusters in knot A. We identify roughly 600 clusters, which is a large amount of clusters for a smaller galaxy. This is primarily due to the lower peaks and thresholds we set; higher thresholds for the stellar peak and minimum density ($100\Msol\,{\rm pc}^{-3}$) resulted in a few hundred clusters. This choice did not affect our conclusions, but helped to identify more of the clusters in knot A.

\begin{figure}
    \includegraphics[width=0.48\textwidth]{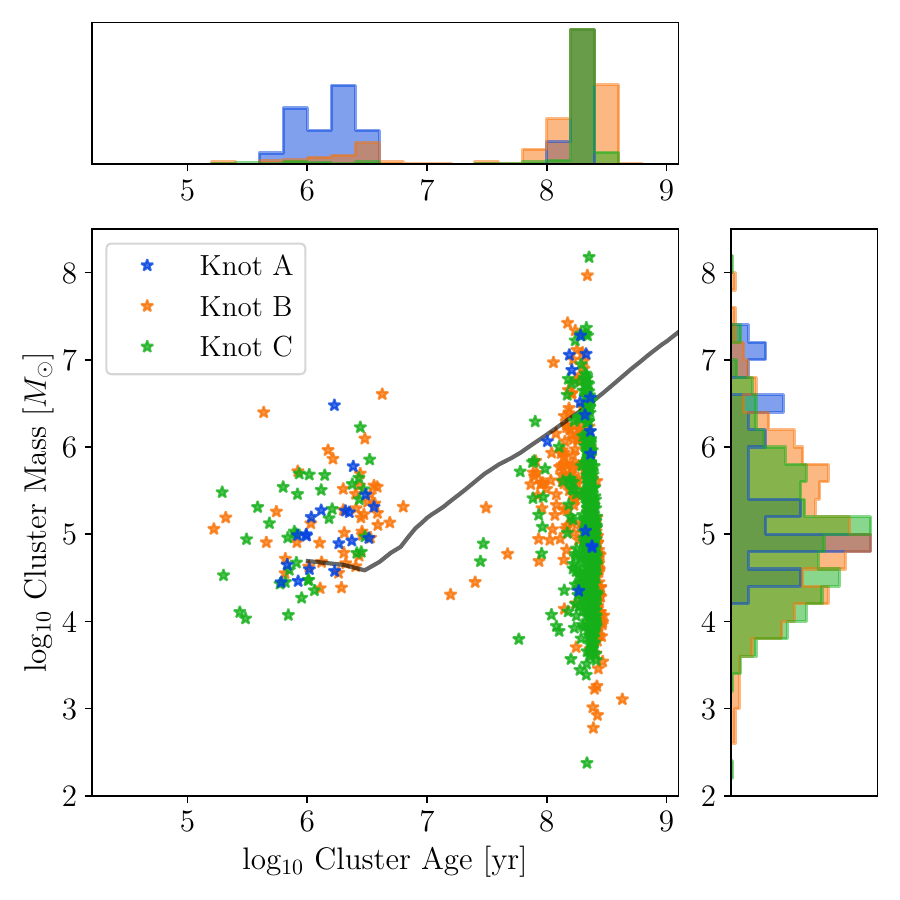}
    \caption{Distribution of masses and ages of the stellar clusters in the three knots, calculated as described in Section~\ref{sec:knot_properties}. The heights of the histograms are normalised to the same maximum height. The black line shows the 90\,\% completeness of the stellar populations in \citet{Sirressi+22}. }
    \label{fig:knot_mass_age}
\end{figure}

The distribution of mass and median ages of the clusters, divided into their knot affiliation, are presented in Figure~\ref{fig:knot_mass_age}. As mentioned in Section~\ref{sec:sfr}, we notice two massive populations of star clusters that form during the starburst peaks that coincide with the close interactions. These groups of population ages, $\sim10^6$\,Myr and $\sim10^8$\,Myr, have been found in stellar population synthesis models of Haro 11 \citep[][]{Chandar+23, PapaderosOstlin2023}. We recover in the simulation star clusters as massive as those found in the observations at the youngest age, while the majority of the old population would remain undetected at masses below a few times $10^5\Msol$; as can be seen from the 90\% completeness limit from \citet[][]{Sirressi+22}, presented as a dark line. The most massive clusters found in our simulations $\sim 10^8\Msol$ are not found in the observations, but could stem from insufficient mass resolution, the cluster identification, or other, more spurious, effects. For future work, Ejdetjärn et al. (in prep), we plan to perform more extensive analysis of the cluster formation within our high-resolution simulations of the same setup.

\subsection{Kinematics}\label{sec:kinematics}
\begin{figure*}
    \includegraphics[width=0.4\textwidth]{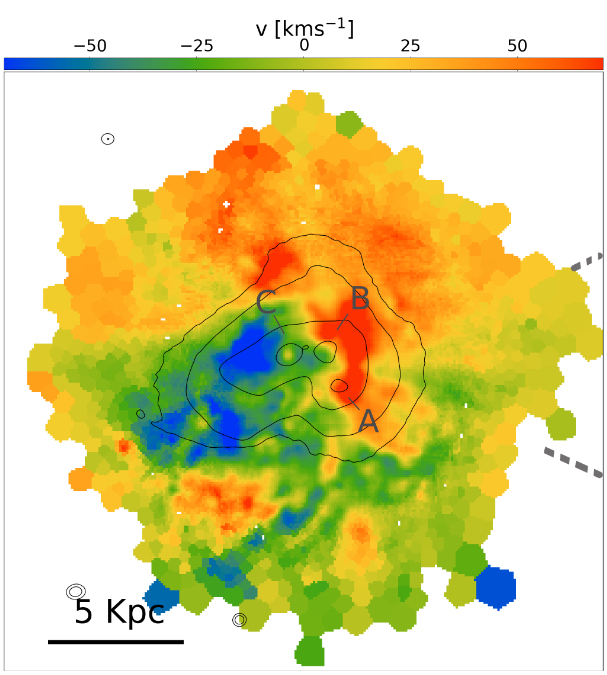}
    \includegraphics[width=0.47\textwidth]{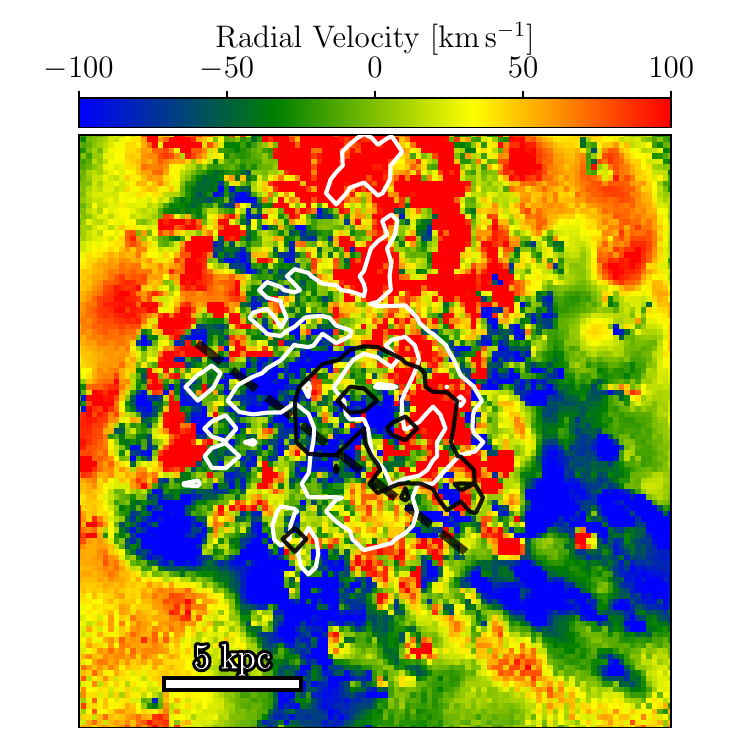}
    \caption{ The \Halpha velocity field of observations in \citet{Menacho+21} (\emph{left}; their Figure 2) and our simulation (\emph{right}). The simulation shows contours of the stellar (black contours) and total gas (white contours) annotated. The black dashed line is used for the analysis of Figure~\ref{fig:pv-diagram}. } 
    \label{fig:vel_maps}
\end{figure*}

In this section we present the gas kinematics of the inner galaxy, while the tidal tail kinematics is presented in Section~\ref{sec:tidal_tail}. We produce mock observations of the \Halpha velocity field, in order to compare with the \Halpha observations from \citet{Menacho+21}. Velocity maps from their observations and our simulation are presented in the left and right plots, respectively, in Figure~\ref{fig:vel_maps}. The mock \Halpha velocity was calculated as the weighted-mean of velocities, using \Halpha emissivities \citep[calculated following][]{Ejdetjarn+22, Ejdetjarn+24} as weights. Additionally, the gas and stellar distribution is annotated as black and white contours, respectively.

The velocity maps show a broad likeness, with the gas being blue-shifted on the left side of the galaxy and the red-shifted on the right side. Additionally, knots A and B are within the gas moving towards us, this kinematic likely originates from the disc arm on this side. Knot C is located between a region of blue- and red-shifted gas, which also agrees with our simulation. However, there is a discrepancy between the magnitude of the velocity maps, as our simulations have about twice the speed of the observations. We discuss possible origins and solutions to this mismatch in Section~\ref{sec:discussion}.

The observations by \citet{Menacho+21} show channels of ionised gas cones, which they identify as outflows from stellar feedback. However, it is not within the scope of this paper to disentangle outflows from the \Halpha velocity field in our simulation, as it would not provide a reliable comparison; the \Halpha velocity field is sensitive to the specifics of stellar formation and feedback recipes, as well as the stochastic nature of feedback. 

\begin{figure*}
    \includegraphics[width=0.84\textwidth]{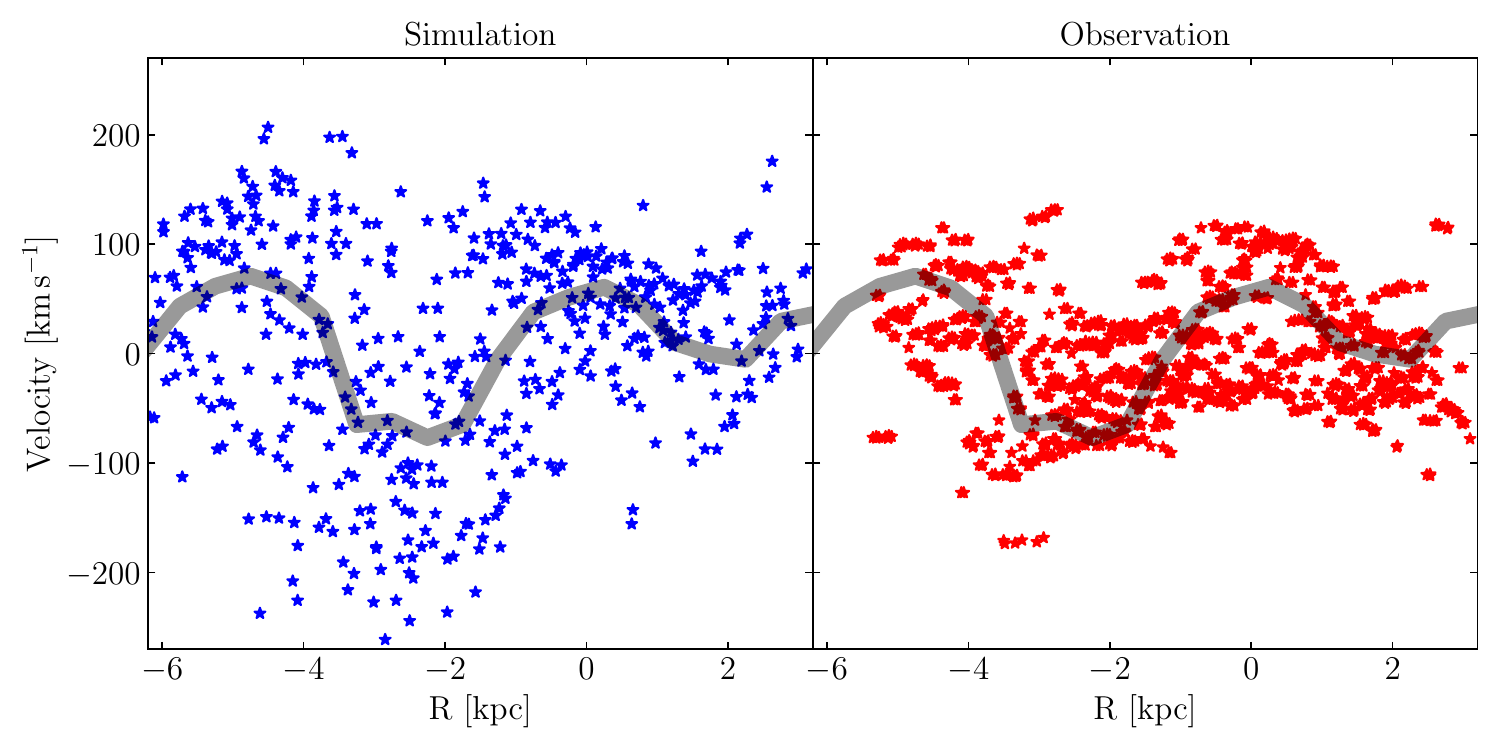}
    \caption{ The PV diagram of our simulations (\emph{left}) and observational data (\emph{right}) of Haro 11. The $x$-axis corresponds to the projected length along the black line in Figure~\ref{fig:vel_maps}. The observational data was reproduced from Ö15, which they obtained using the {\small CIGALE} instrument. The grey curve in both plots is the weighted-mean of the simulation data in distance bins of 0.5 kpc. See Section~\ref{sec:kinematics} for details on how these plots were made. }  
    \label{fig:pv-diagram}
\end{figure*}

In Ö15, the authors made a comparison between the position-velocity (PV) diagram of the Antennae galaxies and Haro 11, which they noted had a similar shape and velocity magnitude. They acquired data using the {\small CIGALE} instrument and archived data for the Antennae. Their data covered a straight line going from knot C to A, and extending a few kpc beyond them. We follow their method and present our own PV diagram in Figure~\ref{fig:pv-diagram}.  The $x$-axis is determined as the projected distance along the dark line on the left plot in Figure~\ref{fig:vel_maps}. The left/right plot in the figure shows the simulation/observational data, which have very similar shapes with peaks and valleys at roughly the same distance intervals. We highlight these similarities, with a broad grey line in both plots, which is the mean velocity of the simulation at each distance in bins of 0.5\,kpc. This figure also highlights that several of the simulations cells attain much higher speeds, as much as twice as fast, as the observations. However, the mean velocity (grey line) follows the same sine-like shape and is within the limits of the observational data.

\section{Additional tests and the impact of specific parameter variations}\label{sec:extra_tests}
We next summarise the impact of some specific parameter changes on the resulting properties of our fiducial Haro 11 model. In particular, the compared properties in the previous section: morphology, star formation, and kinematics. To summarise, we find that these properties are stable to minor variations in individual parameters but the exact morphology and SFR is dependent on the complex interplay between several parameters. Additionally, the parameters that have the most significant impact are those that alter the orbit, e.g. mass ratio and initial velocity.

\subsection{Producing one tidal tail}\label{sec:match_tidal_tail}
\begin{figure*}
    \includegraphics[width=0.90\textwidth]{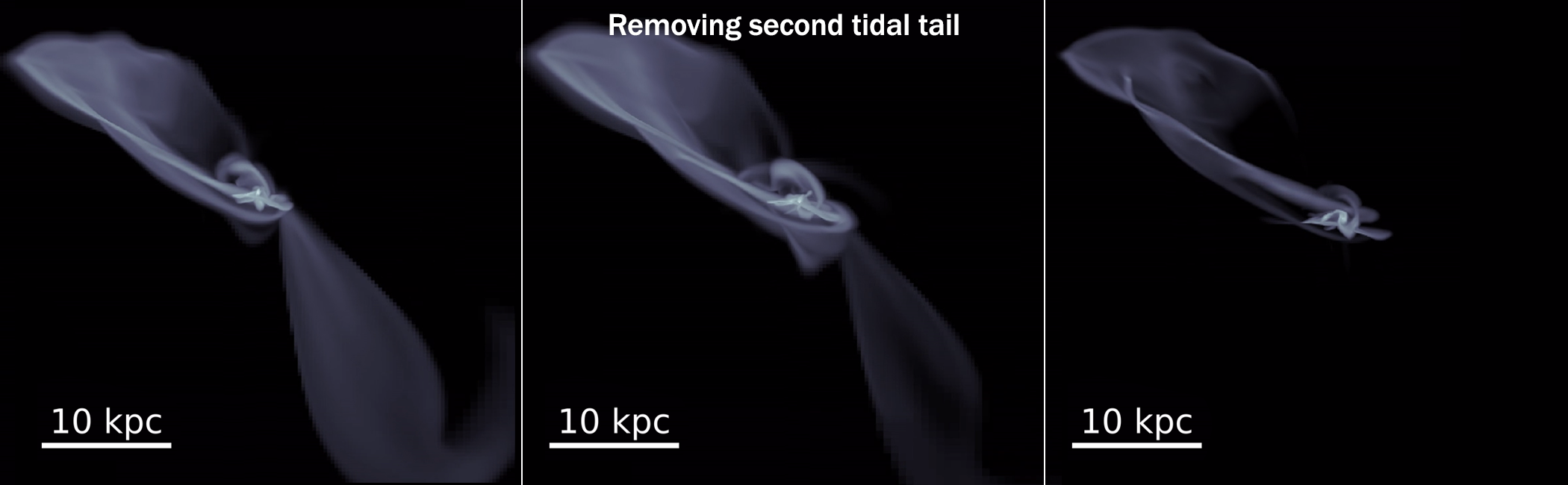}\vspace{0.5mm}
    \includegraphics[width=0.90\textwidth]{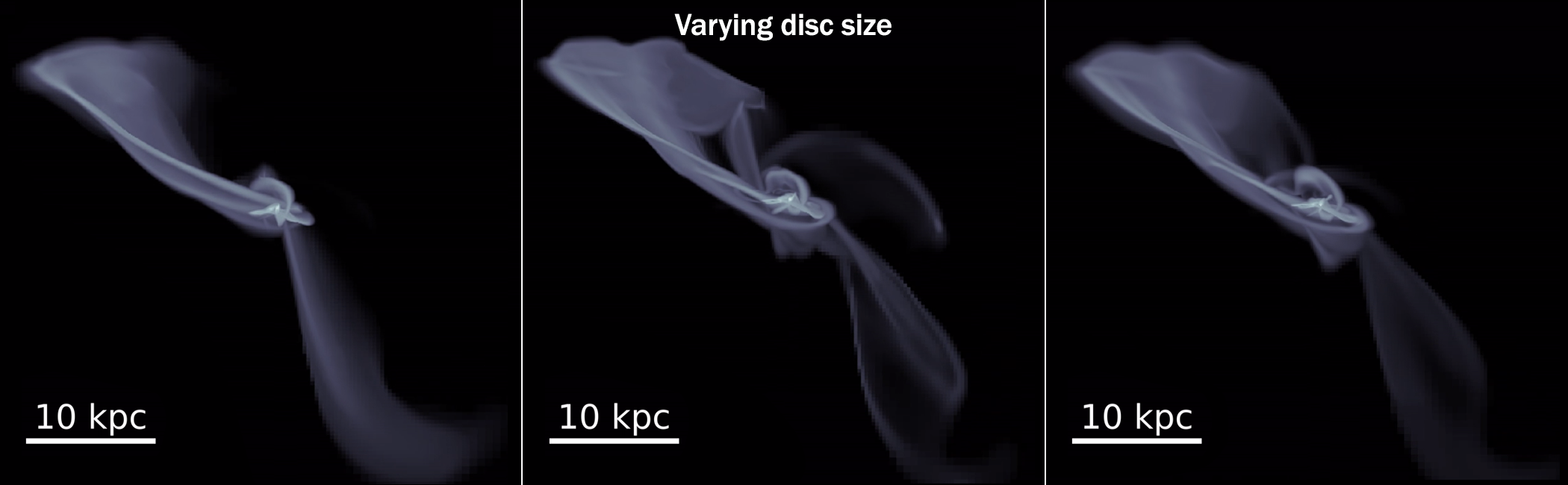}\vspace{0.5mm}
    \hspace*{1mm}\includegraphics[width=0.90\textwidth]{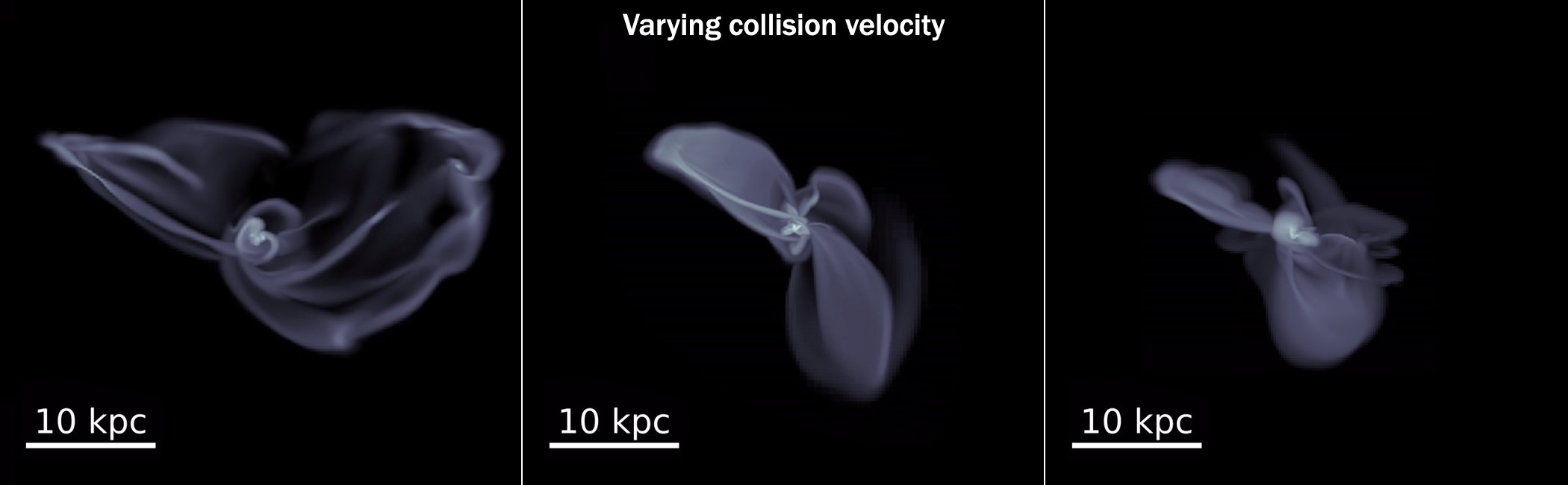}
    \caption{ Projection maps of the gas densities, with each row showing the impact of specific parameters.
    \textbf{\emph{Top}}: Highlights the successive removal of the second tidal tail, from left to right. \emph{Left}: one prograde and one retrograde galaxy. \emph{Middle}: reducing the gas mass and size of the retrograde galaxy. \emph{Right}: changing the mass ratio between the two galaxies, by reducing the dark matter mass of the retrograde galaxy.
    \textbf{\emph{Centre}}: Highlights the morphological impact of disc size. \emph{Left}: One galaxy has retrograde motion and the other has prograde. \emph{Middle}: Thicker disc for the retrograde. \emph{Right}: Truncating the gas and stellar disc at 2/3 the original radius.
    \textbf{\emph{Bottom}}: Highlights the morphological impact of the initial velocity. \emph{Left}: The original Antennae initial conditions (scaled down in size), see Section~\ref{sec:ICs} for details. \emph{Middle}: Slower initial speed, causing a shorter merger time and more direct impact. \emph{Right}: A more direct impact between the galaxies, which forms more debris.}
    \label{fig:tidal_tail_removal}
\end{figure*}

In the previous section we demonstrated that our fiducial simulation is able to reproduce the \emph{one} tidal tail, as observed in Haro 11, along with some of its features. However, producing a single tidal tail in this setup is non-trivial and requires a certain combination of parameters. In particular, in the top row of Figure~\ref{fig:tidal_tail_removal} we illustrate that one of the galaxies need to have a retrograde orbit \emph{and} the merger needs to be asymmetric (difference in mass ratio and/or gas fraction) to fully suppress the retrograde tail. The figure shows three projection maps of the gas density that, from left to right, highlight the decrease in tidal tail strength: from prograde and retrograde galaxies (left), to a smaller and less massive retrograde disc (middle), to larger mass ratio difference between the galaxies (right). Additionally, it can be surmised from the rightmost image that the mass ratio also affects the strength of the tidal tail from the prograde galaxy, as a lower mass companion galaxy is less efficient at stripping gas. 

Additionally, truncating the gas disc also has an effect on the resulting morphology of the tidal tails. In the centre row of Figure~\ref{fig:tidal_tail_removal} we show an initial galaxy with retrograde and prograde motions (left), a thicker disc for the retrograde (middle), and a truncated disc for the retrograde (right). The thicker disc can be seen to yield a slightly weaker second tail and have more tidal features surrounding it. The truncated disc also shows a more diffuse second tail and possibly some small changes to the main tail.

From our simulation tests, we learned that the length of the tidal tail puts a constraint on the time since the first interaction, when the tail first forms. In order for the gas tail to reach the observed 40\,kpc, it would need roughly 200 Myr to expand. For comparison, a shorter interaction time  of $\sim100$\,Myr would yield a length of $\sim$20 kpc. This time might be a lower limit, as the observed expansion speed of tidal tail is slightly lower in the \HI observations. Furthermore, the presence of a tidal tail this long also substantiates that Haro 11 is at least in its second close interaction.

\subsection{Varying velocity} 
We briefly review the effects of varying the initial velocities, as velocity is connected to characteristics of the tidal tail and the stellar populations (the interaction times are connected to the SFR bursts). Generally, increasing the speed of the progenitors extended the time before the second encounter, giving the tidal tail more time to extend further. However, it also leads to a very brief interaction time which results in a weak tidal tail. In contrast, lower velocities lead to a quicker merger and shorter tidal tail, but the gas within the tail is denser.

In the bottom row of Figure~\ref{fig:tidal_tail_removal}, we visualise the impact of changes in speed and angle. The leftmost image shows a prograde and retrograde galaxy at a 1.2 times speed of the fiducial tests (see left plot in centre row for comparison) and the middle image is at a 0.8 times the speed. The right image shows a more direct impact angle, resulting in more gas debris and a shorter interaction time. As an aside, the simulation with lower initial speed did not seem to result in any significant change in the magnitude of the velocity field.

\subsection{Elliptical vs disc}
\begin{figure}
    \centering\includegraphics[width=0.45\textwidth]{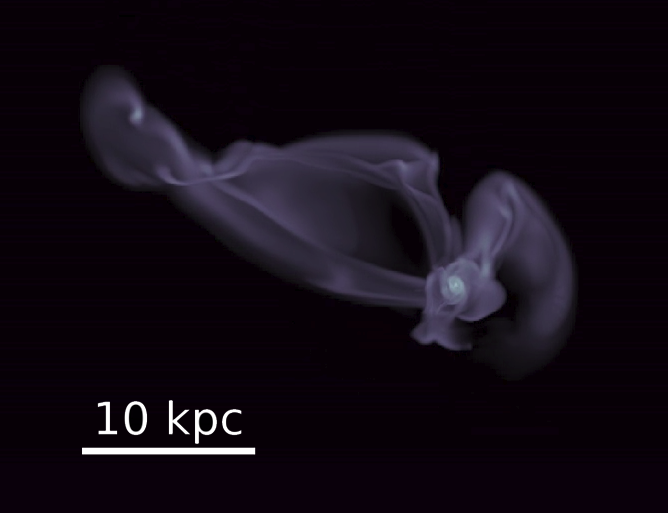}
    \caption{ Projection map of the gas density in a setup with an elliptical galaxy and a prograde galaxy. }
    \label{fig:elliptical}
\end{figure}

A formation scenario proposed mentioned in \citet[][]{Adamo+10} is that the progenitor galaxies of Haro 11 could be a low-mass, evolved galaxy and a gas rich (disc) galaxy, due to the irregular gas morphology around the knots and stellar age distribution (see discussion in Section~\ref{sec:obs_constraint}). We briefly explored a setup where the retrograde disc was replaced by a gas-poor, dwarf elliptical galaxy. For this test, we kept the total mass of the systems the same, but the gas and stars were redistributed between the galaxies to replace the prograde disc with a sphere\footnote{The stellar distribution followed a Hernquist profile with similar scale radius as the original disc.} of stars with no gas. An image of the tidal tail gas density is presented in Figure~\ref{fig:elliptical}, which shows that this approach also exhibits only one tidal tail; as to be expected when there is no disc material to be stripped. Additionally, the extended spherical shape causes some additional tidal substructures to appear. We did not analyse the inner morphology.

The SFR of this scenario is lower leading up to the first encounter, after which it is comparable to other test simulations with discs. Although, it does not reach the same peak in SFR nor exhibit any distinct bump during the first encounter, compared to the fiducial simulation in Figure~\ref{fig:sfr}. This does not necessarily rule out this specific merger scenario, as other changes, such as a higher gas fraction in the prograde galaxy, could be made to boost the SFR. 
\subsection{Density of the CGM}\label{sec:cgm_test}
\begin{figure*}
\includegraphics[width=0.97\textwidth]{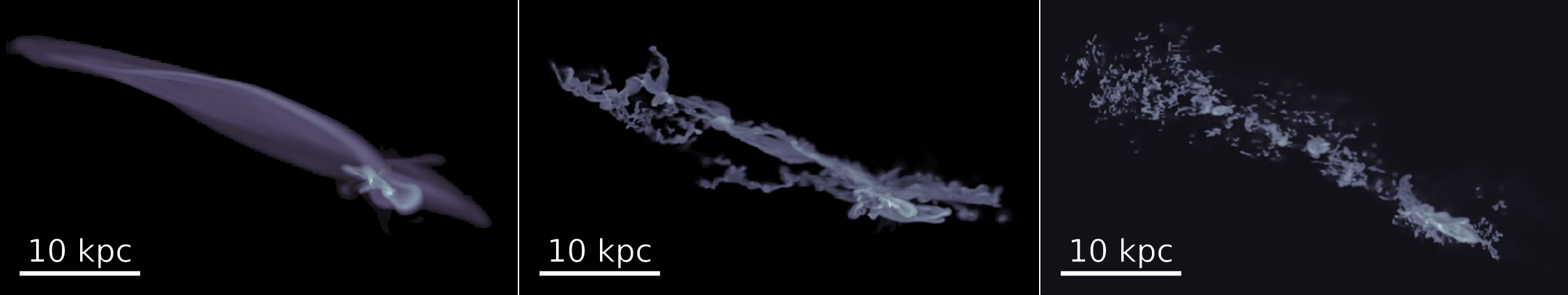}
    \caption{ Density maps of the tidal tail in low-resolution simulations of varying circumgalactic density. From left to right, the densities are: ${\rm log}_{10}\,\rho_{\rm CGM}=(-6, -5, -4)\,{\rm m_H\,cm}^{-3}$. }
    \label{fig:cgm_increase}
\end{figure*}

Additionally, we performed a few tests varying the circumgalactic density. The fiducial value $\rho_{\rm CGM}=10^{-4}\,{\rm m_H\, cm}^{-3}$ is calculated by spreading some of the gas mass in a 60$^3$\,kpc$^3$ cube, i.e. the size of the largest of the two dark matter halos. This gas mass was taken as the baryonic mass missing from the galaxy mass to reach the cosmic baryon fraction. In Figure~\ref{fig:cgm_increase} we show an example of how the increased density affects the tidal tail structure. The tail can be seen to go from being one coherent structure at low density and becomes clumpy towards higher density. Comparably, observations of the \HI tail show that the tail is clumpy in nature \citep[][]{LeReste+24}, which could be due to interactions with a dense halo of gas or clumpiness within disc before or during stripping. Notably, a more gaseous halo would also affect the shock structure of ejected gas and could be compared with observed ionisation structures, but due to these examples having no stellar feedback there is no evident difference in outflow shocks.

\section{Discussion}\label{sec:discussion}

\subsection{Caveats and parameter variation} 
We have in the previous sections demonstrated that our simulation setup of two disc galaxies undergoing a merger is able to reproduce the key features of the morphology, kinematics, SFR, and stellar population of Haro 11. Finding a model that is able to reproduce several different features of a specific galaxy is not trivial and is encouraging for the reliability of the presented merger scenario. However, there are caveats to our results: the model is not accurate enough for detailed comparison and different setups can yield similar results (e.g. morphologies). Thus, it is difficult to attain a definitively best set of parameters. Here, we connect back to the observational data to address the strength and weaknesses of our simulation model. 

\subsubsection{Tidal tail}
The simulation in this work show that the presence of a {\it single} tidal tail, as reported by \citet{LeReste+24}, is in agreement with the scenario that Haro 11 is currently undergoing a merger. This is further confirmed by new observational data presented in this work of a stellar tidal tail, which is the first time the stellar component has been captured. However, because the Haro 11 tidal tail is observed as a 2D projection and does not show any curvature or other distinct morphological features, it is difficult to model the tidal tail morphology in detail, which limits the constraint to the length and width of the tail. This puts more emphasis on also matching the kinematics and other properties, such as the relative \HI gas fraction. Overall, we find that the properties of the observed tail match well with our fiducial Haro 11 simulation, and since the properties of the tidal tail are closely connected to the orbital conditions, this is a good indication that the orbit of the fiducial simulation is correct. Another good indication is our discovery that a specific set of parameters are required for the second tidal tail to not be present/detectable, which limits the allowed setups. 

However, the tidal tail kinematics in our simulation were slightly blue-shifted compared to observations, which we mainly attribute to how the zero-point of the velocity is determined. As our simulation is an isolated volume, there is no clear systemic velocity. Each galaxy in the merger has a different central LOS velocity, which in our simulations gives a range of 'systemic' velocities between -50\kmsec to +0\kmsecalt. Thus, the tidal tail can become more blue- or red-shifted depending on the adopted zero-velocity. Another important factor is the choice of observational angle, as even relatively small changes could shift the tail velocity significantly (but might negatively impact other matches). Additionally, observations show a variation in systemic velocity between different methods and tracers, which could account for the relatively small offset.

\subsubsection{Inner morphology \& kinematics}
The inner morphology of Haro 11 has a few distinct features that we are able to match: the three stellar knots, the disc arm of gas and stars ('ear'), the 'bent' shape of the galaxy, in particular knot A and B being covered by gas. Fitting the exact positions of the knots (relative each other) and the orientation of the disc arm requires detailed tweaking of orbital parameters. Additionally, the relative position of the knots is closely connected to the viewing angle and the chosen time output. We note that by varying these, the position of these morphological features can be improved. However, changing the angles was found to lead to a worse match with other features; primarily the length, position, and kinematics of the tidal tail.

Furthermore, we have shown that knot B and C are the centres of the progenitor galaxies while knot A is part of the stellar disc around knot B, which is in agreement with observations of the morphology (knot B and C being more compact than knot A) and the relative mass between the knots. Currently, the galaxies have had a close interaction and are moving past each other, reaching their apocenter a few Myr later into the merger. Just a few Myr after this, the prograde galaxy's bulge will be closer to where knot A is expected and in knot B's stead a very young stellar population will form from the intense gas compression between the galaxies. Thus, the morphology of the galaxy shifts sharply during the merger, which presents an alternative merger scenario where knot B is not a core but mainly composed of young stars. However, this alternative scenario is unlikely based on the knot's observed compactness and stellar ages. 

From our mock observations in Figure~\ref{fig:mock}, knot A is not visible. Instead, we infer its presence from the stellar mass distribution, young stellar ages, and the placement of the disc arm. In agreement with observations, there is a presence of dust and gas around knot A, but our simulation show more dust along the line of sight which obscures the knot. Additionally, the amount and density of young, luminous stars in knot A could affects its visibility. The formation of luminous clusters depends on the specifics of small-scale physics, such as the star formation efficiency and gas clumping, as well as the initial conditions, e.g. the gas fraction and disc properties. Additionally, the morphology depends on the timing and viewing angle for the analysis.

The velocity field is one of the most challenging components to utilise as a comparison with our simulations, as the velocities are highly sensitive to numerous factors, e.g. the orbits, rotational velocity, stellar feedback, and viewing angle. Primarily, it serves to confirm the general kinematics resulting from the orbital parameters, rather than providing precise constraints. To this end, our simulation matches the general appearance of the observed velocity field. However, the inner \Halpha velocity reaches speeds twice the magnitude of observations, which indicates that the initial orbit or disc rotation should be slowed down. The issue with reducing the orbital speed is that the length and velocity of the tidal tail is sensitive to this parameter, and would decrease its length while also cause more gas to be stripped (making the second tail more visible). Thus, a change in the orbit or disc properties appears as the more likely solution, but would require extensive exploration of a different parameter space. 

\subsubsection{Stellar and gas properties}
The SFR of our chosen output is slightly higher than the observed SFR ($\sim40\Msolyr$ compared to $\lesssim30\Msolyr$). This increases to $\gtrsim100\Msolyr$ within the time it takes the galaxies to merge and the morphology to become unrecognisable. Even if we chose this later output, the factor of two difference in the SFR is well within the uncertainties of small-scale physics within the simulation that regulate star formation, e.g. the adopted star formation efficiency and stellar feedback. We found that the SFR was stable to small tweaks to the orbit, with the exception of close interactions. However, the SFR is highly sensitive to the time output we choose to analyse the galaxy, due to the fact that the Haro 11 model is currently undergoing a powerful starburst which can vary from a few $\Msolyr$ to $100\Msolyr$ within $\sim 10$\,Myr. Thus, the current SFR sets a relatively strong constraint for the time at which we observe the galaxy. 

The stellar clusters in our simulation show similar findings as in observations: knot B and C are more massive and contain older stars, compared to knot A. However, our simulation suggest the formation of more massive clusters, which would be readily observed if real. These massive clusters are partially a side-effect of how the post-processing clustering algorithm works, as we chose parameters that enhanced clustering to highlight low mass clusters. Furthermore, we are not fully capturing all stellar cluster (in particular the least massive) due to the limited spatial and mass resolution in our simulations.

The final gas and stellar mass between simulation and observations are in agreement, but this is neither surprising nor a viable constraint as the galaxy is modelled to have similar masses. However, it is reassuring that no additional effects, such as stellar feedback and tidal stripping, have significantly altered this match. Furthermore, the comparison between gas phases is uncertain, but it is encouraging that there is a broad agreement in the distribution. Although, there is a significant amount ($1.90\times10^9\Msol$) of hot/warm ionised gas within the halo of our galaxies, which is too diffuse to be observed and therefore presented separately from the ionised gas mass. 

To summarise, from exploring the orbital parameter space of this merger, we recognise there are numerous ways to match the features of Haro 11 and, thus, many formation scenarios are possible. However, to find a set of parameters that produce several key aspects of the galaxy is not trivial. This is the first time Haro 11 has ever been simulated with hydrodynamical simulations, and that our simulations match several observed Haro 11 features is thus highly encouraging. Furthermore, it validates the hypothesis that Haro 11 is produced through a merger, and offers a specific possible merger setup. Future work, applying more precise constraints and comparing other aspects of the galaxy, would be helpful to reduce the available parameter space. In particular, in Ejdetjärn et al. (in prep.) we will present results on the LyC and Ly${\rm \alpha}$ properties of this setup to directly compare with observations.

\subsection{Future work and improvements}
The fiducial model and simulation suite presented in this paper show a great match between various properties of Haro 11 and, more widely, offers a basis for modelling the Haro 11 galaxy, and other BCGs. By building and improving on this model, the comparison with observations can be done in more detail, help understand the origin of other features, and possibly allow more quantitative analysis. However, even minimal improvements require substantial work and therefore left for future investigations. Here we highlight a few improvements that could be made.

There are certain properties of the progenitor galaxies that could be tweaked without significantly changing the orbit. For example, the initial bulge masses we use here are almost an order of magnitude higher than what is suggested through dynamical reasoning (Ö15) for the knots. We performed simple tests to check the impact of lowering the mass of the bulge, as well as removing it entirely. We find that the overall SFR remains largely unchanged, but that heavier bulges form more stars towards the centre \citep[see also][]{MihosHernquist1996}, thus it could change the relative mass between stellar populations within the knots. Additionally, there is room to alter the gas masses, as they primarily impact the SFR and density of the tidal tails.

The relative position between the knots could be improved by changing the inclination of the prograde disc (associated with knot B and A), such that the disc arm (which we have shown can represent knot A) falls better in place where knot A is expected. This is a relatively minor change in the parameters, and we have noted that the tidal tail is stable to the small variations required to shift the disc arm by $\sim1$ kpc. However, this is not guaranteed as the inclination of the prograde galaxy is strongly connected to the morphology of the tidal tail and could worsen this match; or even another property, through complex dynamics.

Additionally, we were not able to conclusively exclude that one of the progenitor galaxies (i.e. the retrograde) is an elliptical galaxy. Our tests of this scenario showed that the overall SFR over time would drop, along with the stellar populations within knot C. However, this can be adjusted, if needed, by changing other parameters, such as the prograde gas fraction or orbital parameters. 

Finally, we will explore this same Haro 11 model at a higher spatial resolution and with on-the-fly radiative transfer in Ejdetjärn et al. (in prep.), where we will focus on the escape of LyC and Ly${\rm \alpha}$ radiation and stellar cluster formation during a galaxy merger. We will employ post processing radiative transfer and evaluate the covering fraction to understand the impact of gas stripping and stellar feedback.

\section{Conclusions}\label{sec:conclusion}
In this paper we have explored possible formation scenarios for the Haro 11 galaxy through numerical simulations of mergers between two galaxies. We perform a large suite of $\sim 500$ hydrodynamical simulations by varying parameters that regulate the orbit and galactic properties of two disc galaxies. By using observational data as constraints, we are able to develop a fiducial Haro 11 model that show a broad agreement with several different observed features. Additionally, we present observations of the first direct detection of a stellar tidal tail in Haro 11.

Our main findings are as follows:
\begin{enumerate}
    \item The orbital and galactic parameters of the progenitor galaxies interact in a complex manner that affects the final morphology, SFR, and stellar properties. We have highlighted how specific parameter changes can alter the merger scenario, and thus the final properties, of Haro 11. However, we find that the orbits are stable to small parameter variations and, thus, minor improvements could be made through tweaking the fiducial model further.
    \item To produce the distinct \emph{single} tidal tail observed in Haro 11, one of the progenitor galaxies needs to have a retrograde motion with the orbit. This weakens the second tidal tail, but to render it completely undetectable the galaxies need to be asymmetric; the retrograde needs to be either smaller, have less gas, or be more massive/dense (to overcome the gravitational pull from tidal forces). Other setups could also produce the one tidal tail, but many would likely show a mismatch for other features.

    \item For out fiducial setup, we are able to reproduce the following qualities of Haro 11:
    \begin{itemize}
        \item The length, kinematics, and relative mass fraction of the \HI tidal tail. 
        \item The inner stellar and gaseous morphology. In particular, we show that the three stellar knots in Haro 11 can be found within our simulation.
        \item The \Halpha velocity field of the galaxy shows several similar qualities to the observed \Halpha kinematics, However, the speeds are notably larger than observed. 
        \item The stellar cluster mass and age distributions. However, the ages are more uncertain, as the simulations suggest the observed population $\lesssim 100\,$Myr should be slightly older (to overlap with the starburst from the first interaction). 
    \end{itemize}
    \item Accepting our fiducial model as a possible formation scenario for Haro 11, we are able to hypothesise about the origin of specific observed components:
    \begin{itemize}
        \item The two progenitor galaxies that form Haro 11 are currently undergoing their second close interaction and will fully merge within a few tens of Myr.
        \item In order for the tidal tail to expand to the observed length of 40 kpc, our simulations require $200-250$\,Myr between the first and second interaction.
        \item The inner morphology of the galaxy is made up from a gaseous disc, which is interacting in a prograde motion and having its disc stripped to form a tidal tail.
        \item The prograde galaxy has its centre where we would expect knot B, and its progenitor is a disc galaxy.
        \item An arm from the stellar and gas disc reaches down to form the diffuse stellar clusters that make up knot A. This arm also makes up the 'ear' that goes from knot A to the right of knot B.
        \item Knot C contains mainly old stars ($\sim100$\,Myr) and its progenitor is likely a gas-poor galaxy.
        \item The galaxy centres will merge and become indistinguishable within $10-20$\,Myr, creating another brief starburst period.
    \end{itemize}
\end{enumerate}

In summary, our simulations demonstrate that a two-disc galaxy merger can replicate the observed features of Haro 11 (as hypothesised in Ö15), including its tidal tail, starburst activity, and stellar morphology. This supports the hypothesis that mergers play a central role in the formation of compact, starburst galaxies like Haro 11. While degeneracies remain in the parameter space, our results provide a strong foundation for future studies exploring the detailed mechanisms driving star formation and LyC escape in merging systems. Advancing this work with higher-resolution models and more precise observational constraints will further refine our understanding of galaxy evolution in extreme environments.

\section*{Acknowledgements}
The computations and data handling were enabled by resources provided by the National Academic Infrastructure for Supercomputing in Sweden (NAISS), partially funded by the Swedish Research Council through grant agreement no. 2022-06725. We made use of the medium computation allocations and small storage allocation. 

OA acknowledges support from the Knut and Alice Wallenberg Foundation, the Swedish Research Council (grant 2019-04659), and the Swedish National Space Agency (SNSA Dnr 2023-00164).

FR acknowledges support provided by the University of Strasbourg Institute for Advanced Study (USIAS), within the French national programme Investment for the Future (Excellence Initiative) IdEx-Unistra. 

The {\small MeerKAT} telescope is operated by the South African Radio Astronomy Observatory, which is a facility of the National Research Foundation, an agency of the Department of Science and Innovation. 

For our analysis we have used {\small PYTHON}, including packages: numpy \citep[][]{Harris+20}, matplotlib for {\small PYTHON} \citep[][]{Hunter+07}. Loading and analysis of the simulation data was done using tools from the yt-project \citep[][]{Turk+11}.

\section*{Data Availability}
The data underlying this article will be shared on reasonable request to the corresponding author.
 



\bibliographystyle{mnras}
\input{mnras.bbl} 




\appendix


\section{\HI velocity map of the absorption}\label{app:tidal_tail_abs_map}
\begin{figure}
    \includegraphics[width=0.46\textwidth]{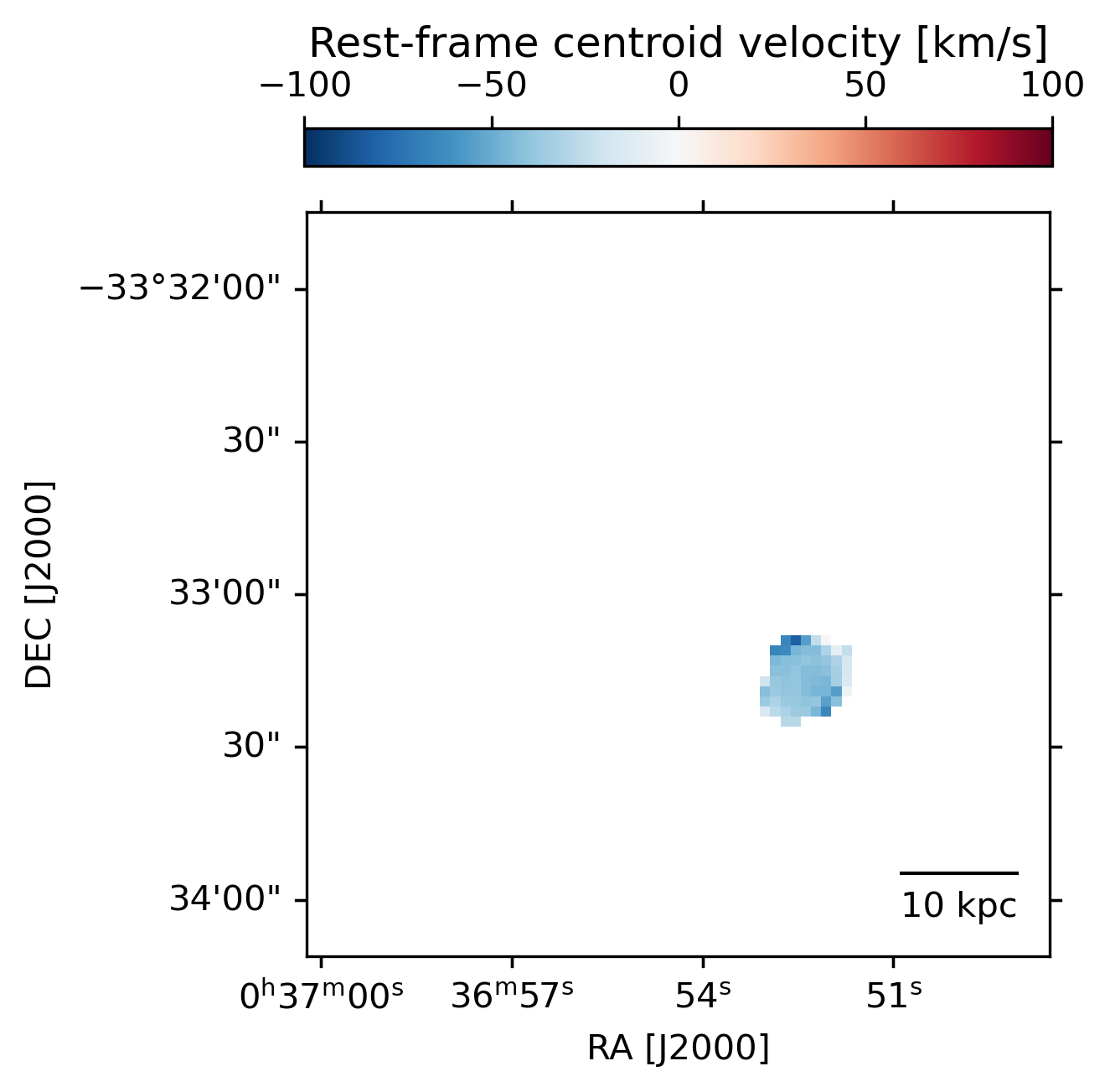}
    \caption{  Moment-1 (velocity centroid) map of the 21-cm line in absorption from {\small MeerKAT} observations \citep[][]{LeReste+24}.  } 
    \label{fig:tidal_tail_vel_emit_absorb}
\end{figure}

The velocity map of neutral \HI emission presented in Figure~\ref{fig:tidal_tail_velocity} shows emission in both the inner parts and tidal tail. Meanwhile, the absorption presented here in Figure~\ref{fig:tidal_tail_vel_emit_absorb} shows significant absorption only in the inner parts of the galaxy. Furthermore, it indicates that there is a large outflow of neutral gas at velocities $\sim50$\kmsec towards us.

\section{Optical observations of stellar tidal tail}\label{app:stellar_tidal_tail}
The detection of the stellar tidal tail in Haro 11, as discussed and shown in Section~\ref{sec:tidal_tail}, was done with VLT/FORS (program 113.26LV). Observations were done for a net integration time of 6240s in the B-band (filter b\_HIGH+113 split in 8 exposures of 780s) and 5760s in I (I-BESS+77, split in 24 exposures of 240s). The shorter exposure time in I-band was chosen to ensure that the sky background did not saturate the detector. The observations were split in 4 observing blocks, 2 in each filter. The B-band observations were obtained in Dark (fractional lunar illumination, FLI , 30\%) time and the I-band in grey time (FLI $\sim$40\%). The observations were performed in service mode, under clear conditions and with a seeing from 1 to 1.5 arcsec. The observations used the Standard Resolution (SR) collimator and the red sensitive (MIT) CCD with 2x2 binning offering a pixel scale of 0.25"/pix.

The images were processed using standard techniques, using the ESO/MIDAS software, and using standard calibration data obtained on the same night(s). We first created a master bias which was subtracted from all science and other calibration frames. We then created a normalised flatfield using the twilight calibration frames. A cosmic ray filtering (FILTER/COSMIC) was applied to the science data which were then flat-fielded, corrected for airmass, CCD gain and normalised to 1 s. Residual image artefacts (bad columns and charge bleeding trails) were flagged. Residual sky emission was removed from each frame by fitting a first order polynomial surface (using the routine FIT/FLAT). Finally, all images for each filter were registered and combined. We note that the centre of each knot in Haro 11, as well as some relatively bright neighbouring stars, are saturated in the FORS images. As we here focus on the extended low surface brightness emission around Haro 11, this is not a major concern. For photometric zero-point we used the values for the observing nights from the FORS photometric trending analysis, available on the following ESO webpage:
\url{https://eso.org/observing/dfo/quality/FORS2/reports/HEALTH/trend\_report\_ZEROPOINTS\_BH\_HC.html}

We also verified the calibration for an unsaturated star next to Haro 11 that is also detected in the HST/ACS/WFC/F435W image, available on MAST.


\bsp	
\label{lastpage}
\end{document}